\ifx\newheadisloaded\relax\immediate\write16{***already loaded}\endinput\else\let\newheadisloaded=\relax\fi
\gdef\islinuxolivetti{F}
\gdef\PSfonts{T}
\magnification\magstep1

\newdimen\papwidth
\newdimen\papheight
\newskip\beforesectionskipamount  
\newskip\sectionskipamount 
\def\sectionskip{\vskip\sectionskipamount}
\def\beforesectionskip{\vskip\beforesectionskipamount}
\papwidth=16truecm
\if F\islinuxolivetti
\papheight=22truecm
\voffset=0.4truecm
\hoffset=0.4truecm
\else
\papheight=16truecm
\voffset=-1.5truecm
\hoffset=0.4truecm
\fi
\hsize=\papwidth
\vsize=\papheight
\nopagenumbers
\headline={\ifnum\pageno>1 {\hss\tenrm-\ \folio\ -\hss} \else
{\hfill}\fi}
\newdimen\texpscorrection
\texpscorrection=0.15truecm 

\def\sectionsize{\twelvepoint}
\def\sectiontype{\bf}
\def\subsectionsize{}
\def\subsectiontype{\bf}
\def\em{\sl}
\newfam\truecmsy
\newfam\truecmr
\newfam\msbfam
\newfam\scriptfam
\newfam\truecmsy
\newskip\ttglue 
\if T\islinuxolivetti
\papheight=11.5truecm
\fi
\if F\PSfonts
\font\twelverm=cmr12
\font\tenrm=cmr10
\font\eightrm=cmr8
\font\sevenrm=cmr7
\font\sixrm=cmr6
\font\fiverm=cmr5

\font\twelvebf=cmbx12
\font\tenbf=cmbx10
\font\eightbf=cmbx8
\font\sevenbf=cmbx7
\font\sixbf=cmbx6
\font\fivebf=cmbx5

\font\twelveit=cmti12
\font\tenit=cmti10
\font\eightit=cmti8
\font\sevenit=cmti7
\font\sixit=cmti6
\font\fiveit=cmti5

\font\twelvesl=cmsl12
\font\tensl=cmsl10
\font\eightsl=cmsl8
\font\sevensl=cmsl7
\font\sixsl=cmsl6
\font\fivesl=cmsl5

\font\twelvei=cmmi12
\font\teni=cmmi10
\font\eighti=cmmi8
\font\seveni=cmmi7
\font\sixi=cmmi6
\font\fivei=cmmi5

\font\twelvesy=cmsy10	at	12pt
\font\tensy=cmsy10
\font\eightsy=cmsy8
\font\sevensy=cmsy7
\font\sixsy=cmsy6
\font\fivesy=cmsy5
\font\twelvetruecmsy=cmsy10	at	12pt
\font\tentruecmsy=cmsy10
\font\eighttruecmsy=cmsy8
\font\seventruecmsy=cmsy7
\font\sixtruecmsy=cmsy6
\font\fivetruecmsy=cmsy5

\font\twelvetruecmr=cmr12
\font\tentruecmr=cmr10
\font\eighttruecmr=cmr8
\font\seventruecmr=cmr7
\font\sixtruecmr=cmr6
\font\fivetruecmr=cmr5

\font\twelvebf=cmbx12
\font\tenbf=cmbx10
\font\eightbf=cmbx8
\font\sevenbf=cmbx7
\font\sixbf=cmbx6
\font\fivebf=cmbx5

\font\twelvett=cmtt12
\font\tentt=cmtt10
\font\eighttt=cmtt8

\font\twelveex=cmex10	at	12pt
\font\tenex=cmex10

\font\twelvemsb=msbm10	at	12pt
\font\tenmsb=msbm10
\font\eightmsb=msbm8
\font\sevenmsb=msbm7
\font\sixmsb=msbm6
\font\fivemsb=msbm5

\font\twelvescr=eusm10 at 12pt
\font\tenscr=eusm10
\font\eightscr=eusm8
\font\sevenscr=eusm7
\font\sixscr=eusm6
\font\fivescr=eusm5
\fi
\if T\PSfonts
\font\twelverm=ptmr	at	12pt
\font\tenrm=ptmr	at	10pt
\font\eightrm=ptmr	at	8pt
\font\sevenrm=ptmr	at	7pt
\font\sixrm=ptmr	at	6pt
\font\fiverm=ptmr	at	5pt

\font\twelvebf=ptmb	at	12pt
\font\tenbf=ptmb	at	10pt
\font\eightbf=ptmb	at	8pt
\font\sevenbf=ptmb	at	7pt
\font\sixbf=ptmb	at	6pt
\font\fivebf=ptmb	at	5pt

\font\twelveit=ptmri	at	12pt
\font\tenit=ptmri	at	10pt
\font\eightit=ptmri	at	8pt
\font\sevenit=ptmri	at	7pt
\font\sixit=ptmri	at	6pt
\font\fiveit=ptmri	at	5pt

\font\twelvesl=ptmro	at	12pt
\font\tensl=ptmro	at	10pt
\font\eightsl=ptmro	at	8pt
\font\sevensl=ptmro	at	7pt
\font\sixsl=ptmro	at	6pt
\font\fivesl=ptmro	at	5pt

\font\twelvei=cmmi12
\font\teni=cmmi10
\font\eighti=cmmi8
\font\seveni=cmmi7
\font\sixi=cmmi6
\font\fivei=cmmi5

\font\twelvesy=cmsy10	at	12pt
\font\tensy=cmsy10
\font\eightsy=cmsy8
\font\sevensy=cmsy7
\font\sixsy=cmsy6
\font\fivesy=cmsy5
\font\twelvetruecmsy=cmsy10	at	12pt
\font\tentruecmsy=cmsy10
\font\eighttruecmsy=cmsy8
\font\seventruecmsy=cmsy7
\font\sixtruecmsy=cmsy6
\font\fivetruecmsy=cmsy5

\font\twelvetruecmr=cmr12
\font\tentruecmr=cmr10
\font\eighttruecmr=cmr8
\font\seventruecmr=cmr7
\font\sixtruecmr=cmr6
\font\fivetruecmr=cmr5

\font\twelvebf=cmbx12
\font\tenbf=cmbx10
\font\eightbf=cmbx8
\font\sevenbf=cmbx7
\font\sixbf=cmbx6
\font\fivebf=cmbx5

\font\twelvett=cmtt12
\font\tentt=cmtt10
\font\eighttt=cmtt8

\font\twelveex=cmex10	at	12pt
\font\tenex=cmex10

\font\twelvemsb=msbm10	at	12pt
\font\tenmsb=msbm10
\font\eightmsb=msbm8
\font\sevenmsb=msbm7
\font\sixmsb=msbm6
\font\fivemsb=msbm5

\font\twelvescr=eusm10 at 12pt
\font\tenscr=eusm10
\font\eightscr=eusm8
\font\sevenscr=eusm7
\font\sixscr=eusm6
\font\fivescr=eusm5
\fi
\def\eightpoint{\def\rm{\fam0\eightrm}%
\textfont0=\eightrm
  \scriptfont0=\sixrm
  \scriptscriptfont0=\fiverm 
\textfont1=\eighti
  \scriptfont1=\sixi
  \scriptscriptfont1=\fivei 
\textfont2=\eightsy
  \scriptfont2=\sixsy
  \scriptscriptfont2=\fivesy 
\textfont3=\tenex
  \scriptfont3=\tenex
  \scriptscriptfont3=\tenex 
\textfont\itfam=\eightit
  \scriptfont\itfam=\sixit
  \scriptscriptfont\itfam=\fiveit 
  \def\it{\fam\itfam\eightit}%
\textfont\slfam=\eightsl
  \scriptfont\slfam=\sixsl
  \scriptscriptfont\slfam=\fivesl 
  \def\sl{\fam\slfam\eightsl}%
\textfont\ttfam=\eighttt
  \def\tt{\fam\ttfam\eighttt}%
\textfont\bffam=\eightbf
  \scriptfont\bffam=\sixbf
  \scriptscriptfont\bffam=\fivebf
  \def\bf{\fam\bffam\eightbf}%
\textfont\scriptfam=\eightscr
  \scriptfont\scriptfam=\sixscr
  \scriptscriptfont\scriptfam=\fivescr
  \def\script{\fam\scriptfam\eightscr}%
\textfont\msbfam=\eightmsb
  \scriptfont\msbfam=\sixmsb
  \scriptscriptfont\msbfam=\fivemsb
  \def\bb{\fam\msbfam\eightmsb}%
\textfont\truecmr=\eighttruecmr
  \scriptfont\truecmr=\sixtruecmr
  \scriptscriptfont\truecmr=\fivetruecmr
  \def\truerm{\fam\truecmr\eighttruecmr}%
\textfont\truecmsy=\eighttruecmsy
  \scriptfont\truecmsy=\sixtruecmsy
  \scriptscriptfont\truecmsy=\fivetruecmsy
\tt \ttglue=.5em plus.25em minus.15em 
\normalbaselineskip=9pt
\setbox\strutbox=\hbox{\vrule height7pt depth2pt width0pt}%
\normalbaselines
\rm
}

\def\tenpoint{\def\rm{\fam0\tenrm}%
\textfont0=\tenrm
  \scriptfont0=\sevenrm
  \scriptscriptfont0=\fiverm 
\textfont1=\teni
  \scriptfont1=\seveni
  \scriptscriptfont1=\fivei 
\textfont2=\tensy
  \scriptfont2=\sevensy
  \scriptscriptfont2=\fivesy 
\textfont3=\tenex
  \scriptfont3=\tenex
  \scriptscriptfont3=\tenex 
\textfont\itfam=\tenit
  \scriptfont\itfam=\sevenit
  \scriptscriptfont\itfam=\fiveit 
  \def\it{\fam\itfam\tenit}%
\textfont\slfam=\tensl
  \scriptfont\slfam=\sevensl
  \scriptscriptfont\slfam=\fivesl 
  \def\sl{\fam\slfam\tensl}%
\textfont\ttfam=\tentt
  \def\tt{\fam\ttfam\tentt}%
\textfont\bffam=\tenbf
  \scriptfont\bffam=\sevenbf
  \scriptscriptfont\bffam=\fivebf
  \def\bf{\fam\bffam\tenbf}%
\textfont\scriptfam=\tenscr
  \scriptfont\scriptfam=\sevenscr
  \scriptscriptfont\scriptfam=\fivescr
  \def\script{\fam\scriptfam\tenscr}%
\textfont\msbfam=\tenmsb
  \scriptfont\msbfam=\sevenmsb
  \scriptscriptfont\msbfam=\fivemsb
  \def\bb{\fam\msbfam\tenmsb}%
\textfont\truecmr=\tentruecmr
  \scriptfont\truecmr=\seventruecmr
  \scriptscriptfont\truecmr=\fivetruecmr
  \def\truerm{\fam\truecmr\tentruecmr}%
\textfont\truecmsy=\tentruecmsy
  \scriptfont\truecmsy=\seventruecmsy
  \scriptscriptfont\truecmsy=\fivetruecmsy
\tt \ttglue=.5em plus.25em minus.15em 
\normalbaselineskip=12pt
\setbox\strutbox=\hbox{\vrule height8.5pt depth3.5pt width0pt}%
\normalbaselines
\rm
}

\def\twelvepoint{\def\rm{\fam0\twelverm}%
\textfont0=\twelverm
  \scriptfont0=\tenrm
  \scriptscriptfont0=\eightrm 
\textfont1=\twelvei
  \scriptfont1=\teni
  \scriptscriptfont1=\eighti 
\textfont2=\twelvesy
  \scriptfont2=\tensy
  \scriptscriptfont2=\eightsy 
\textfont3=\twelveex
  \scriptfont3=\twelveex
  \scriptscriptfont3=\twelveex 
\textfont\itfam=\twelveit
  \scriptfont\itfam=\tenit
  \scriptscriptfont\itfam=\eightit 
  \def\it{\fam\itfam\twelveit}%
\textfont\slfam=\twelvesl
  \scriptfont\slfam=\tensl
  \scriptscriptfont\slfam=\eightsl 
  \def\sl{\fam\slfam\twelvesl}%
\textfont\ttfam=\twelvett
  \def\tt{\fam\ttfam\twelvett}%
\textfont\bffam=\twelvebf
  \scriptfont\bffam=\tenbf
  \scriptscriptfont\bffam=\eightbf
  \def\bf{\fam\bffam\twelvebf}%
\textfont\scriptfam=\twelvescr
  \scriptfont\scriptfam=\tenscr
  \scriptscriptfont\scriptfam=\eightscr
  \def\script{\fam\scriptfam\twelvescr}%
\textfont\msbfam=\twelvemsb
  \scriptfont\msbfam=\tenmsb
  \scriptscriptfont\msbfam=\eightmsb
  \def\bb{\fam\msbfam\twelvemsb}%
\textfont\truecmr=\twelvetruecmr
  \scriptfont\truecmr=\tentruecmr
  \scriptscriptfont\truecmr=\eighttruecmr
  \def\truerm{\fam\truecmr\twelvetruecmr}%
\textfont\truecmsy=\twelvetruecmsy
  \scriptfont\truecmsy=\tentruecmsy
  \scriptscriptfont\truecmsy=\eighttruecmsy
\tt \ttglue=.5em plus.25em minus.15em 
\setbox\strutbox=\hbox{\vrule height7pt depth2pt width0pt}%
\normalbaselineskip=15pt
\normalbaselines
\rm
}
%
\fontdimen16\tensy=2.7pt
\fontdimen13\tensy=4.3pt
\fontdimen17\tensy=2.7pt
\fontdimen14\tensy=4.3pt
\fontdimen18\tensy=4.3pt
\fontdimen16\eightsy=2.7pt
\fontdimen13\eightsy=4.3pt
\fontdimen17\eightsy=2.7pt
\fontdimen14\eightsy=4.3pt
\fontdimen18\sevensy=4.3pt
\fontdimen16\sevensy=1.8pt
\fontdimen13\sevensy=4.3pt
\fontdimen17\sevensy=2.7pt
\fontdimen14\sevensy=4.3pt
\fontdimen18\sevensy=4.3pt
%
\def\hexnumber#1{\ifcase#1 0\or1\or2\or3\or4\or5\or6\or7\or8\or9\or
 A\or B\or C\or D\or E\or F\fi}
\mathcode`\=="3\hexnumber\truecmr3D
\mathchardef\not="3\hexnumber\truecmsy36
\mathcode`\+="2\hexnumber\truecmr2B
\mathcode`\(="4\hexnumber\truecmr28
\mathcode`\)="5\hexnumber\truecmr29
\mathcode`\!="5\hexnumber\truecmr21
\mathcode`\(="4\hexnumber\truecmr28
\mathcode`\)="5\hexnumber\truecmr29

\def\ddot{\mathaccent"0\hexnumber\truecmr7F }
\def\tilde{\mathaccent"0\hexnumber\truecmr7E }

\def\hat{\mathaccent"0\hexnumber\truecmr5E }
\def\dot{\mathaccent"0\hexnumber\truecmr5F }
\def\Phi{\mathchar"0\hexnumber\truecmr08 }
\def\Gamma {\mathchar"0\hexnumber\truecmr00 }
\def\Delta {\mathchar"0\hexnumber\truecmr01 }
\def\Theta {\mathchar"0\hexnumber\truecmr02 }
\def\Lambda{\mathchar"0\hexnumber\truecmr03 }
\def\Xi {\mathchar"0\hexnumber\truecmr04 }
\def\Pi{\mathchar"0\hexnumber\truecmr05 }
\def\Sigma{\mathchar"0\hexnumber\truecmr06 }
\def\Upsilon {\mathchar"0\hexnumber\truecmr07 }
\def\Phi {\mathchar"0\hexnumber\truecmr08 }
\def\Psi {\mathchar"0\hexnumber\truecmr09 }
\def\Omega{\mathchar"0\hexnumber\truecmr0A }
\newcount\EQNcount \EQNcount=1
\newcount\CLAIMcount \CLAIMcount=1
\newcount\SECTIONcount \SECTIONcount=0
\newcount\SUBSECTIONcount \SUBSECTIONcount=1
\def\ifff(#1,#2,#3){\ifundefined{#1#2}%
\expandafter\xdef\csname #1#2\endcsname{#3}\else%
\fi}
\def\NEWDEF #1,#2,#3 {\ifff({#1},{#2},{#3})}
\def\actualnumber{\number\SECTIONcount}
\def\EQ(#1){\lmargin(#1)\eqno\tag(#1)}
\def\NR(#1){&\lmargin(#1)\tag(#1)\cr}  
\def\tag(#1){\lmargin(#1)({\rm \actualnumber}.\number\EQNcount)
 \NEWDEF e,#1,(\actualnumber.\number\EQNcount)
\global\advance\EQNcount by 1
}
\def\SECT(#1)#2\par{\lmargin(#1)\SECTION#2\par
\NEWDEF s,#1,{\actualnumber}\noindent
}
\def\SUBSECT(#1)#2\par{\lmargin(#1)
\SUBSECTION#2\par 
\NEWDEF s,#1,{\actualnumber.\number\SUBSECTIONcount}
}
\def\CLAIM #1(#2) #3\par{
\vskip.1in\medbreak\noindent
{\lmargin(#2)\bf #1\ \actualnumber.\number\CLAIMcount.} {\sl #3}\par
\NEWDEF c,#2,{#1\ \actualnumber.\number\CLAIMcount}
\global\advance\CLAIMcount by 1
\ifdim\lastskip<\medskipamount
\removelastskip\penalty55\medskip\fi}
\def\CLAIMNONR #1(#2) #3\par{
\vskip.1in\medbreak\noindent
{\lmargin(#2)\bf #1.} {\sl #3}\par
\NEWDEF c,#2,{#1}
\global\advance\CLAIMcount by 1
\ifdim\lastskip<\medskipamount
\removelastskip\penalty55\medskip\fi}
\def\SECTION#1\par{\vskip0pt plus.3\vsize\penalty-75
    \vskip0pt plus -.3\vsize
    \global\advance\SECTIONcount by 1
    \beforesectionskip\noindent
{\sectionsize\sectiontype \actualnumber.\ #1}
    \EQNcount=1
    \CLAIMcount=1
    \SUBSECTIONcount=1
    \nobreak\sectionskip\noindent}
\def\SECTIONNONR#1\par{\vskip0pt plus.3\vsize\penalty-75
    \vskip0pt plus -.3\vsize
    \global\advance\SECTIONcount by 1
    \beforesectionskip\noindent
{\sectionsize\sectiontype  #1}
     \EQNcount=1
     \CLAIMcount=1
     \SUBSECTIONcount=1
     \nobreak\sectionskip\noindent}
\def\SUBSECTION#1\par{\vskip0pt plus.2\vsize\penalty-75%
    \vskip0pt plus -.2\vsize%
    \beforesectionskip\noindent%
{\subsectionsize\subsectiontype \actualnumber.\number\SUBSECTIONcount.\ #1}
    \global\advance\SUBSECTIONcount by 1
    \nobreak\sectionskip\noindent}
\def\SUBSECTIONNONR#1\par{\vskip0pt plus.2\vsize\penalty-75
    \vskip0pt plus -.2\vsize
\beforesectionskip\noindent
{\subsectionsize\subsectiontype #1}
    \nobreak\sectionskip\noindent\noindent}
\def\ifundefined#1{\expandafter\ifx\csname#1\endcsname\relax}
\def\equ(#1){\ifundefined{e#1}$\spadesuit$#1\else\csname e#1\endcsname\fi}
\def\clm(#1){\ifundefined{c#1}$\spadesuit$#1\else\csname c#1\endcsname\fi}
\def\sec(#1){\ifundefined{s#1}$\spadesuit$#1
\else Section \csname s#1\endcsname\fi}
\let\endarg=\par
\def\finish{\def\endarg{\par\endgroup}}
\def\start{\endarg\begingroup}

 \def\beginFROM{\start\parskip=0pt\vskip\baselineskip
\def\finish{\def\endarg{\egroup\par\endgroup}}
  \vbox\bgroup\obeylines\eightpoint\em\finish}

\def\ABSTRACT#1\par{
\vskip 1in {\noindent\sectionsize\sectiontype Abstract.} #1 \par}

\def\TODAY{\number\day~\ifcase\month\or January \or February \or March \or
April \or May \or June
\or July \or August \or September \or October \or November \or December \fi
\number\year\timecount=\number\time
\divide\timecount by 60
}
\newcount\timecount
\def\DRAFT{\def\lmargin(##1){\strut\vadjust{\kern-\strutdepth
\vtop to \strutdepth{
\baselineskip\strutdepth\vss\rlap{\kern-1.2 truecm\eightpoint{##1}}}}}
\font\footfont=cmti7
\footline={{\footfont \hfil File:\jobname, \TODAY,  \number\timecount h}}
}
\newbox\strutboxJPE
\setbox\strutboxJPE=\hbox{\strut}
\def\subitem#1#2\par{\vskip\baselineskip\vskip-\ht\strutboxJPE{\item{#1}#2}}
\gdef\strutdepth{\dp\strutbox}
\def\lmargin(#1){}
\def\period{\unskip.\spacefactor3000 { }}
%
%
\newbox\noboxJPE
\newbox\byboxJPE
\newbox\paperboxJPE
\newbox\yrboxJPE
\newbox\jourboxJPE
\newbox\pagesboxJPE
\newbox\volboxJPE
\newbox\preprintboxJPE
\newbox\toappearboxJPE
\newbox\bookboxJPE
\newbox\bybookboxJPE
\newbox\publisherboxJPE
\newbox\inprintboxJPE
\def\refclearJPE{
   \setbox\noboxJPE=\null             \gdef\isnoJPE{F}
   \setbox\byboxJPE=\null             \gdef\isbyJPE{F}
   \setbox\paperboxJPE=\null          \gdef\ispaperJPE{F}
   \setbox\yrboxJPE=\null             \gdef\isyrJPE{F}
   \setbox\jourboxJPE=\null           \gdef\isjourJPE{F}
   \setbox\pagesboxJPE=\null          \gdef\ispagesJPE{F}
   \setbox\volboxJPE=\null            \gdef\isvolJPE{F}
   \setbox\preprintboxJPE=\null       \gdef\ispreprintJPE{F}
   \setbox\toappearboxJPE=\null       \gdef\istoappearJPE{F}
   \setbox\inprintboxJPE=\null        \gdef\isinprintJPE{F}
   \setbox\bookboxJPE=\null           \gdef\isbookJPE{F}  \gdef\isinbookJPE{F}
     
   \setbox\bybookboxJPE=\null         \gdef\isbybookJPE{F}
   \setbox\publisherboxJPE=\null      \gdef\ispublisherJPE{F}
     
}

\def\ref{\refclearJPE\bgroup}
\def\no   {\egroup\gdef\isnoJPE{T}\setbox\noboxJPE=\hbox\bgroup}
\def\by   {\egroup\gdef\isbyJPE{T}\setbox\byboxJPE=\hbox\bgroup}
\def\paper{\egroup\gdef\ispaperJPE{T}\setbox\paperboxJPE=\hbox\bgroup}
\def\yr{\egroup\gdef\isyrJPE{T}\setbox\yrboxJPE=\hbox\bgroup}
\def\jour{\egroup\gdef\isjourJPE{T}\setbox\jourboxJPE=\hbox\bgroup}
\def\pages{\egroup\gdef\ispagesJPE{T}\setbox\pagesboxJPE=\hbox\bgroup}
\def\vol{\egroup\gdef\isvolJPE{T}\setbox\volboxJPE=\hbox\bgroup\bf}
\def\preprint{\egroup\gdef
\ispreprintJPE{T}\setbox\preprintboxJPE=\hbox\bgroup}
\def\toappear{\egroup\gdef
\istoappearJPE{T}\setbox\toappearboxJPE=\hbox\bgroup}
\def\inprint{\egroup\gdef
\isinprintJPE{T}\setbox\inprintboxJPE=\hbox\bgroup}
\def\book{\egroup\gdef\isbookJPE{T}\setbox\bookboxJPE=\hbox\bgroup\em}
\def\publisher{\egroup\gdef
\ispublisherJPE{T}\setbox\publisherboxJPE=\hbox\bgroup}
\def\inbook{\egroup\gdef\isinbookJPE{T}\setbox\bookboxJPE=\hbox\bgroup\em}
\def\bybook{\egroup\gdef\isbybookJPE{T}\setbox\bybookboxJPE=\hbox\bgroup}
\newdimen\refindent
\refindent=5em
\def\endref{\egroup \sfcode`.=1000
 \if T\isnoJPE
 \hangindent\refindent\hangafter=1
      \noindent\hbox to\refindent{[\unhbox\noboxJPE\unskip]\hss}\ignorespaces
     \else  \noindent    \fi
 \if T\isbyJPE    \unhbox\byboxJPE\unskip: \fi
 \if T\ispaperJPE \unhbox\paperboxJPE\unskip\period \fi
 \if T\isbookJPE {\it\unhbox\bookboxJPE\unskip}\if T\ispublisherJPE, \else.
\fi\fi
 \if T\isinbookJPE In {\it\unhbox\bookboxJPE\unskip}\if T\isbybookJPE,
\else\period \fi\fi
 \if T\isbybookJPE  (\unhbox\bybookboxJPE\unskip)\period \fi
 \if T\ispublisherJPE \unhbox\publisherboxJPE\unskip \if T\isjourJPE, \else\if
T\isyrJPE \  \else\period \fi\fi\fi
 \if T\istoappearJPE (To appear)\period \fi
 \if T\ispreprintJPE Pre\-print\period \fi
 \if T\isjourJPE    \unhbox\jourboxJPE\unskip\ \fi
 \if T\isvolJPE     \unhbox\volboxJPE\unskip\if T\ispagesJPE, \else\ \fi\fi
 \if T\ispagesJPE   \unhbox\pagesboxJPE\unskip\  \fi
 \if T\isyrJPE      (\unhbox\yrboxJPE\unskip)\period \fi
 \if T\isinprintJPE (in print)\period \fi
\filbreak
}
\def\hexnumber#1{\ifcase#1 0\or1\or2\or3\or4\or5\or6\or7\or8\or9\or
 A\or B\or C\or D\or E\or F\fi}
\textfont\msbfam=\tenmsb
\scriptfont\msbfam=\sevenmsb
\scriptscriptfont\msbfam=\fivemsb
\mathchardef\varkappa="0\hexnumber\msbfam7B
\newcount\FIGUREcount \FIGUREcount=0
\newdimen\figcenter
\def\fig(#1){\ifundefined{fig#1}%
\global\advance\FIGUREcount by 1%
\NEWDEF fig,#1,{Fig.\ \number\FIGUREcount}
\immediate\write16{ FIG \number\FIGUREcount : #1}
\fi
\csname fig#1\endcsname\relax}
\def\figure #1 #2 #3 #4\cr{\null%
\ifundefined{fig#1}%
\global\advance\FIGUREcount by 1%
\NEWDEF fig,#1,{Fig.\ \number\FIGUREcount}
\immediate\write16{  FIG \number\FIGUREcount : #1}
\fi
{\goodbreak\figcenter=\hsize\relax
\advance\figcenter by -#3truecm
\divide\figcenter by 2
\midinsert\vskip #2truecm\noindent\hskip\figcenter
\includegraphics{#1}\vskip 0.8truecm\noindent \vbox{\eightpoint\noindent
{\bf\fig(#1)}: #4}\endinsert}}
\def\figurewithtex #1 #2 #3 #4 #5\cr{\null%
\ifundefined{fig#1}%
\global\advance\FIGUREcount by 1%
\NEWDEF fig,#1,{Fig.\ \number\FIGUREcount}
\immediate\write16{ FIG \number\FIGUREcount: #1}
\fi
{\goodbreak\figcenter=\hsize\relax
\advance\figcenter by -#4truecm
\divide\figcenter by 2
\midinsert\vskip #3truecm\noindent\hskip\figcenter
\includegraphics{#1}{\hskip\texpscorrection\input #2 }\vskip 0.8truecm\noindent \vbox{\eightpoint\noindent
{\bf\fig(#1)}: #5}\endinsert}}
\def\figurewithtexplus #1 #2 #3 #4 #5 #6\cr{\null%
\ifundefined{fig#1}%
\global\advance\FIGUREcount by 1%
\NEWDEF fig,#1,{Fig.\ \number\FIGUREcount}
\immediate\write16{ FIG \number\FIGUREcount: #1}
\fi
{\goodbreak\figcenter=\hsize\relax
\advance\figcenter by -#4truecm
\divide\figcenter by 2
\midinsert\vskip #3truecm\noindent\hskip\figcenter
\includegraphics{#1}{\hskip\texpscorrection\input #2 }\vskip #5truecm\noindent \vbox{\eightpoint\noindent
{\bf\fig(#1)}: #6}\endinsert}}
\catcode`@=11
\def\footnote#1{\let\@sf\empty 
  \ifhmode\edef\@sf{\spacefactor\the\spacefactor}\/\fi
  #1\@sf\vfootnote{#1}}
\def\vfootnote#1{\insert\footins\bgroup\eightpoint
  \interlinepenalty\interfootnotelinepenalty
  \splittopskip\ht\strutbox 
  \splitmaxdepth\dp\strutbox \floatingpenalty\@MM
  \leftskip\z@skip \rightskip\z@skip \spaceskip\z@skip \xspaceskip\z@skip
  \textindent{#1}\footstrut\futurelet\next\fo@t}
\def\fo@t{\ifcat\bgroup\noexpand\next \let\next\f@@t
  \else\let\next\f@t\fi \next}
\def\f@@t{\bgroup\aftergroup\@foot\let\next}
\def\f@t#1{#1\@foot}
\def\@foot{\strut\egroup}
\def\footstrut{\vbox to\splittopskip{}}
\skip\footins=\bigskipamount 
\count\footins=1000 
\dimen\footins=8in 
\catcode`@=12 

\def\CC{{\script C}}

\def\HH{{\script H}}

\def\HALF{{\textstyle{1\over 2}}}

\def\QED{\hfill\smallskip
         \line{$\hfill{\vcenter{\vbox{\hrule height 0.2pt
	\hbox{\vrule width 0.2pt height 1.8ex \kern 1.8ex
		\vrule width 0.2pt}
	\hrule height 0.2pt}}}$
               \ \ \ \ \ \ }
         \bigskip}
\def\real{{\bf R}}

\def\PROOF{\medskip\noindent{\bf Proof.\ }}
\def\REMARK{\medskip\noindent{\bf Remark.\ }}
\def\LIKEREMARK#1{\medskip\noindent{\bf #1.\ }}
\tenpoint
\normalbaselineskip=5.25mm
\baselineskip=5.25mm
\parskip=10pt
\beforesectionskipamount=24pt plus8pt minus8pt
\sectionskipamount=3pt plus1pt minus1pt
\def\em{\it}

\expandafter\xdef\csname
ehambath\endcsname{(2.1)}
\expandafter\xdef\csname
eeqmo1\endcsname{(2.2)}
\expandafter\xdef\csname
eeqmo3\endcsname{(2.3)}
\expandafter\xdef\csname
cn1\endcsname{Remark 2.1}
\expandafter\xdef\csname
cn2\endcsname{Remark 2.2}
\expandafter\xdef\csname
egibbs\endcsname{(2.4)}
\expandafter\xdef\csname
egdef0\endcsname{(2.5)}
\expandafter\xdef\csname
eeffham\endcsname{(2.6)}
\expandafter\xdef\csname
esteq\endcsname{(3.1)}
\expandafter\xdef\csname
ecteq\endcsname{(3.2)}
\expandafter\xdef\csname
ebigP\endcsname{(3.3)}
\expandafter\xdef\csname
eallxt\endcsname{(3.4)}
\expandafter\xdef\csname
cstrovarku\endcsname{Theorem\ 3.1}
\expandafter\xdef\csname
cchc\endcsname{Theorem\ 3.2}
\expandafter\xdef\csname
cnow\endcsname{Remark 3.3}
\expandafter\xdef\csname
cs1a\endcsname{Remark 3.4}
\expandafter\xdef\csname
ef0\endcsname{(3.5)}
\expandafter\xdef\csname
ef1\endcsname{(3.6)}
\expandafter\xdef\csname
efj\endcsname{(3.7)}
\expandafter\xdef\csname
efn\endcsname{(3.8)}
\expandafter\xdef\csname
efn1\endcsname{(3.9)}
\expandafter\xdef\csname
eDequ\endcsname{(3.10)}
\expandafter\xdef\csname
eDeltaequ\endcsname{(3.11)}
\expandafter\xdef\csname
eMagicFormula\endcsname{(3.12)}
\expandafter\xdef\csname
eControlPb\endcsname{(3.13)}
\expandafter\xdef\csname
cs2\endcsname{Remark 3.5}
\expandafter\xdef\csname
cunique++\endcsname{Theorem\ 3.6}
\expandafter\xdef\csname
cDecay\endcsname{Remark 3.7}
\expandafter\xdef\csname
canalyticity\endcsname{Remark 3.8}
\expandafter\xdef\csname
eeqmo3a\endcsname{(4.1)}
\expandafter\xdef\csname
eVeff\endcsname{(4.2)}
\expandafter\xdef\csname
egdef\endcsname{(4.3)}
\expandafter\xdef\csname
efinal\endcsname{(4.4)}
\expandafter\xdef\csname
eldef\endcsname{(4.5)}
\expandafter\xdef\csname
eltdef\endcsname{(4.6)}
\expandafter\xdef\csname
cr4\endcsname{Remark 4.1}
\expandafter\xdef\csname
ephii\endcsname{(4.7)}
\expandafter\xdef\csname
eentropy\endcsname{(4.8)}
\expandafter\xdef\csname
ephidef\endcsname{(4.9)}
\expandafter\xdef\csname
erdef\endcsname{(4.10)}
\expandafter\xdef\csname
elstar\endcsname{(4.11)}
\expandafter\xdef\csname
eLS\endcsname{(4.12)}
\expandafter\xdef\csname
eLS0\endcsname{(4.13)}
\expandafter\xdef\csname
cLS\endcsname{Theorem\ 4.2}
\expandafter\xdef\csname
cDBC\endcsname{Remark 4.3}
\expandafter\xdef\csname
cLS1\endcsname{Remark 4.4}
\expandafter\xdef\csname
esymm1\endcsname{(4.14)}
\expandafter\xdef\csname
esymm\endcsname{(4.15)}
\expandafter\xdef\csname
elphi\endcsname{(4.16)}
\expandafter\xdef\csname
elstarphi\endcsname{(4.17)}
\expandafter\xdef\csname
crelations\endcsname{Proposition\ 4.5}
\expandafter\xdef\csname
eident1\endcsname{(4.18)}
\expandafter\xdef\csname
centropy\endcsname{Theorem\ 4.6}
\expandafter\xdef\csname
cpositive\endcsname{Remark 4.7}
\expandafter\xdef\csname
eid1\endcsname{(4.19)}
\expandafter\xdef\csname
eLS1\endcsname{(4.20)}
\expandafter\xdef\csname
euuu\endcsname{(4.21)}
\expandafter\xdef\csname
eid2\endcsname{(4.22)}
\expandafter\xdef\csname
ei3\endcsname{(4.23)}
\expandafter\xdef\csname
ei4\endcsname{(4.24)}
\expandafter\xdef\csname
ei5\endcsname{(4.25)}
\expandafter\xdef\csname
ei23\endcsname{(4.26)}
\expandafter\xdef\csname
essj\endcsname{(4.27)}
\expandafter\xdef\csname
eLSj\endcsname{(4.28)}
\expandafter\xdef\csname
ee1\endcsname{(4.29)}
\expandafter\xdef\csname
ephidef3\endcsname{(4.30)}
\expandafter\xdef\csname
ee2\endcsname{(4.31)}
\expandafter\xdef\csname
evarentropy\endcsname{(4.32)}
\expandafter\xdef\csname
eident2\endcsname{(4.33)}
\expandafter\xdef\csname
cpositive1\endcsname{Theorem\ 4.8}
\expandafter\xdef\csname
esys1\endcsname{(4.34)}
\expandafter\xdef\csname
esys2\endcsname{(4.35)}
\expandafter\xdef\csname
esys3\endcsname{(4.36)}
\expandafter\xdef\csname
cflux\endcsname{Corollary\ 4.9}
\def\L{{\rm L}}
\def\R{{\rm R}}
\def\S{{\rm S}}
\def\B{{\rm B}}

\def\hidekappa#1{}
\def\d{{\rm d}}
\def\dt{{\rm d}t}
\def\smm{\sum_{m=1}^M}
\def\Lm{{\L,m}}
\def\Rm{{\R,m}}
\def\frac#1#2{{#1\over #2}}
\def\eff{{\rm eff}}
\def\T{{\rm T}}
\def\ie{{\it i.e.}}
\def\eg{{\it e.g.}}
\def\Tr{{\rm Tr~}}
\def\WW{{\cal W}}
\def\FF{{\cal F}}
\def\UU{{\cal U}}
\def\SS{{\cal S}}
\def\D{{\cal M}}
\def\dpqr{\d r\,\d q\,\d p}
\def\ddpqr{\d r,\d q,\d p}
\def\dspq{\d s\,\d q\,\d p}

\def\dpq{\d q\,\d p}
\def\ddpq{\d q,\d p}
\def\supp{{\rm supp~}}
\def\Q(#1,#2){#1^{(#2)}}
\def\QP(#1,#2){\tilde #1^{(#2)}}
\def\Qp(#1){\tilde #1}
\def\REMARK(#1){\lmargin(#1)
\NEWDEF c,#1,{Remark \actualnumber.\number\CLAIMcount}
\medskip\noindent{\bf Remark \actualnumber.\number\CLAIMcount.~}%
\global\advance\CLAIMcount by 1}
\let\truett=\tt
\fontdimen3\tentt=2pt\fontdimen4\tentt=2pt
\def\tt{\hfill\break\null\kern -3truecm\truett
\#\#\#\#\#\#\#\#\#\#\#\#\#\#  }
\let\epsilon=\varepsilon
\normalbaselineskip=12pt
\baselineskip=12pt
\parskip=0pt
\parindent=22.222pt
\beforesectionskipamount=24pt plus0pt minus6pt
\sectionskipamount=7pt plus3pt minus0pt
\overfullrule=0pt
\hfuzz=2pt
\nopagenumbers
\headline={\ifnum\pageno>1 {\hss\tenrm-\ \folio\ -\hss} \else
{\hfill}\fi}
\if T\islinuxolivetti
\font\titlefont=cmbx10 scaled\magstep2

\font\toplinefont=cmr10
\font\pagenumberfont=cmr10
\let\tenpoint=\rm
\else
\font\titlefont=ptmb at 14 pt

\font\toplinefont=cmcsc10
\font\pagenumberfont=ptmb at 10pt
\fi
\newdimen\itemindent\itemindent=1.5em

\def\textindent#1{\indent\llap{#1\enspace}\ignorespaces}
\def\item{\par\noindent
\hangindent\itemindent\hangafter=1\relax
\setitemmark}
\def\setitemindent#1{\setbox0=\hbox{\ignorespaces#1\unskip\enspace}%
\itemindent=\wd0\relax
\message{|\string\setitemindent: Mark width modified to hold
         |`\string#1' plus an \string\enspace\space gap. }%
}
\def\setitemmark#1{\checkitemmark{#1}%
\hbox to\itemindent{\hss#1\enspace}\ignorespaces}
\def\checkitemmark#1{\setbox0=\hbox{\enspace#1}%
\ifdim\wd0>\itemindent
   \message{|\string\item: Your mark `\string#1' is too wide. }%
\fi}
\def\SECTION#1\par{\vskip0pt plus.2\vsize\penalty-75
    \vskip0pt plus -.2\vsize
    \global\advance\SECTIONcount by 1
    \beforesectionskip\noindent
{\sectionsize\sectiontype \actualnumber.\ #1}
    \EQNcount=1
    \CLAIMcount=1
    \SUBSECTIONcount=1
    \nobreak\sectionskip\noindent}

\setitemindent{\bf H3)}
{\titlefont{\centerline{Entropy Production in Non-Linear, Thermally
Driven }}}
{\titlefont{\centerline {Hamiltonian Systems}}}
\vskip 0.5truecm
{\it{\centerline{Jean-Pierre Eckmann${}^{1,2}$, Claude-Alain
Pillet${}^{3,4}$, Luc Rey-Bellet${}^{1,}${\footnote{${}^{*}$}{\rm Current
address: Dept.~of Mathematics and Physics, Rutgers University, Hill
Center,  Rutgers University, Piscataway NJ 08903, USA}}}}
{\eightpoint
\centerline{${}^1$Section de Math\'ematiques, Universit\'e de Gen\`eve,
CH-1211 Gen\`eve 4, Switzerland}
\centerline{${}^2$Dept.~de Physique Th\'eorique, Universit\'e de
Gen\`eve, CH-1211 Gen\`eve 4, Switzerland}
\centerline{${}^3$PHYMAT, Universit\'e de Toulon,
F-83957 La Garde Cedex, France}
\centerline{${}^4$CPT-CNRS Luminy, F-13288 Marseille Cedex 09, France}}}
\vskip 0.5truecm\headline
{\ifnum\pageno>1 {\toplinefont Entropy Production}
\hfill{\pagenumberfont\folio}\fi}
\vskip 3cm
{\eightpoint\narrower\baselineskip 11pt
\LIKEREMARK{Abstract}We consider a finite chain of non-linear
oscillators coupled at its
ends to two infinite heat baths which are at different temperatures. Using
our earlier results about the existence of a stationary state, we
show rigorously that for arbitrary temperature differences and arbitrary
couplings,
such a system has
a unique stationary state. (This extends our earlier
results for small temperature differences.)
In all these cases, any initial state will converge
(at an unknown rate) to the stationary state.
We
show that this stationary state continually produces entropy. 
The rate
of
entropy production is strictly negative when the temperatures are
unequal
and is proportional to the mean energy flux through the system.

}
\SECTION Introduction

In a recent paper, [EPR], we have studied the existence of a
stationary regime in a non-linear non-equilibrium setup. The model
considered was that of a chain of $n$ non-linear oscillators coupled
at each of its two ends to heat baths which are infinite systems at
two different temperatures. Under suitable conditions which we sketch
below, it has been shown that
for sufficiently small temperature differences between the baths, the
complete system has a {\em unique} stationary state, and that every
initial state converges to it. Of course, this stationary state is
not an equilibrium state but a steady state in which supposedly heat flows
(on average) from the hot bath to the cold one. The aim of this paper
is to show first that this result extends to arbitrary temperature
differences
and that the heat flux across the chain is positive.

It should be noted that this is {\em not} a perturbative result.
To prove the existence and uniqueness of the steady state for
non-linear, boundary driven problems with arbitrary temperature
difference is a difficult problem.
See [GLP] and [GKI] for similar results for a gas of particles in a box
with thermostatting boundary conditions. For other boundary driven
models, see [FGS] (a 1-dimensional hard-core gas).

Our model, whose study was started in [EPR],
combines several desirable features, while still allowing for a
rather complete set of rigorous results.
The main features are the property of being fully Hamiltonian (as those
studied in [FGS] and [SL]), with non-linear interactions, and
a realistic implementation of the retro-action
of the chain on the heat baths.
In particular, the system is self-regulating and we do
not need any Gaussian thermostats 
[PH, EM, H, CELS, GC1, GC2, G].

The reader should
note that in our
model the energy of the chain fluctuates wildly in time and there is
no external dissipation term which prevents the energy of the chain
from diverging to infinity. The baths can exchange energy with the
chain.
Also, since the potentials are not
monotone, several stationary non-equilibrium states could possibly
exist, each corresponding for example to one of the extrema of the
potential.
We show here that on the contrary, there is exactly one stationary
state, no matter how large the temperature difference of the baths is.

Once the uniqueness of the steady state is established, we show that,
away from equilibrium, there is a stationary, strictly positive heat flow 
through the chain and the (thermodynamic) entropy production is
strictly negative. We also discuss briefly (heuristically) a
suitable version of the Gallavotti-Cohen fluctuation relation [ECM,
GC1, GC2, G, K, 
LS] for the entropy production in the context of our model.

\SECTION Set-up and Notations

To make this paper accessible without the necessity of referring back
to [EPR], we introduce again the model.
It deals with an anharmonic chain driven at its
ends by two heat baths.

The chain consists of $n$ particles moving in $\real^d$, with $n$ arbitrary
but finite,
and its dynamics is described by the following Hamiltonian:
$$
H_{\S}( q_1, \dots, q_n,p_1, \dots, p_n)\,=\, \sum_{j=1}^{n}
\frac{p_j^2}{2} + V(q_1, \dots, q_n)~,
$$
where the potential $V$ is of the form:\footnote{${}^1$}{The two-body
potential is slightly more restrictive than in
[EPR], since we only take functions of the coordinate differences.}

$$
V(q)\,=\,\sum_{j=1}^n U_j^{(1)}(q_j)
+ \sum_{i=1}^{n-1} U^{(2)}_{i}(q_i-q_{i+1}) ~.
$$
We make the following assumptions on the potential $V$:
\item{\bf H1)}Behavior at infinity: We assume that $V$ is of the form
$$
V(q)\,=\,\HALF\bigl (q-a,{\bf Q}(q-a)\bigr )+F(q)~,
$$
where ${\bf Q}$ is a positive definite ($dn\times dn$) matrix, $a$ is a
vector, and $\partial_{q_i^{(\nu)}}F\in\FF$ for $i=1,\dots,n$ and
$\nu=1,\dots,d$.
Here, $\FF$ denotes the space of those $\CC^\infty $ functions $F$ on
$\real^{dn}$ 
for which
$
\partial^\alpha F(q)
$
is bounded  uniformly in $q\in\real^{dn}$,
for all multi-indices $\alpha $.
\item{\bf H2)}Coupling: Each of the ($d\times d$) matrices
$$
\D_{i}(x)\,\equiv\,{\rm D}^2 U^{(2)}_i(x)~,\quad i=1,\dots,n-1~,
$$
of second derivatives, is
either uniformly positive or negative definite for $x\in \real^d$.

Each heat bath is modeled by an infinite dimensional linear
Hamiltonian system, which is a scalar field whose dynamics is
governed by a $d$-dimensional wave
equation:
$$
H_{\B}(\phi_{i}, \pi_i) \,=\, \frac{1}{2} \int \d x \bigl( |\nabla
\phi_i|^2 + |\pi_i|^2 \bigr) ~.
\EQ(hambath)
$$
We will denote the heat baths by the subscripts
$\L$ and $\R$ respectively.
We couple the $\L$ heat bath to the first particle of the chain and the
$\R$ heat bath to the $n$-th particle of the chain. We choose a
coupling which is linear both in the field variables and in the
particle variables. The total Hamiltonian of the system is then given by
$$
\eqalign{
H(q,p, \phi_\L,\pi_\L,\phi_\R,\pi_\R)\,&=\, \sum_{i\in \{\L,\R\}}
H_{\B}(\phi_{i}, \pi_i) + H_{\S}(q,p)\cr
\,& +\, q_1 \cdot \int \d x \nabla
\phi_\L(x) \rho_\L(x) +  q_n \cdot \int \d x \nabla
\phi_\R(x) \rho_\R(x) ~. \cr
}
\EQ(eqmo1)
$$
We consider the heat baths at positive temperatures $T_\L$ and $T_\R$
respectively, \ie, we will assume that the initial conditions of the
heat baths are distributed according to the Gaussian measure with mean
zero and covariance $(\cdot,\cdot)_iT_i$, where
$(\cdot,\cdot)_i$ is the scalar product defined by the quadratic form
\equ(hambath), $i\in\{\L,\R\}$.

The following reduction to (essentially only) the variables of the
small system is explained in detail in [EPR]:
We integrate out the variables of the baths and project
the dynamics on the variables of the chain. This leads to
integro-differential stochastic equations.
Under suitable assumptions on the coupling functions $\rho_\L$,
$\rho_\R$, they can be expressed
as Markovian equations upon introducing auxiliary variables $r_{i,m}$
with $i\in\{\L,\R\}$, and $m=1,\dots,M$.
At the end, one obtains (see [EPR] for
details)
the following
system of stochastic differential equations:
$$\eqalign{
\d q_j (t) \,&=\, p_j (t)\dt~, \qquad\qquad\qquad j =1,\dots,n ~,\cr
\d p_1 (t) \,&=\, -\nabla_{q_1}V(q(t))\dt  +\smm r_\Lm(t)\dt~,\cr
\d p_j (t) \,&=\, -\nabla_{q_j }V(q(t))\dt ~,\qquad j =2,\dots,n-1 {}~,\cr
\d p_n (t) \,&=\, -\nabla_{q_n}V(q(t))\dt  +\smm  r_\Rm (t)\dt ~,\cr
\d r_\Lm (t) \,&=\, -\gamma_\Lm  r_\Lm(t)\dt + \lambda_\Lm^2
\gamma_\Lm q_1(t) \dt
                 - \lambda_\Lm\sqrt{{2\gamma_\Lm}{T_\L}}\, \d w_\Lm(t)  ~,\cr
\d r_\Rm (t) \,&=\, -\gamma_\Rm  r_\Rm (t)\dt + \lambda_\Rm^2
\gamma_\Rm q_n(t) \dt
                 - \lambda_\Rm \sqrt{{2\gamma_\Rm}{T_\R}}\, \d
                 w_\Rm(t)~,  \cr
&\quad\quad\quad\quad m=1,\dots,M~,
}\EQ(eqmo3)
$$
which defines a Markov diffusion process on $\real^{2d(n+M)}$.
Each $w_\Lm$ and $w_\Rm$ is a standard $d$-dimensional Brownian motion.

\REMARK(n1) In Eqs.\equ(eqmo3) the variables $r_{\Lm}$, $r_\Rm$
describe both the (random) forces exerted by the heat bath on the chain
and the (dissipative) forces due to the retroaction of the heat baths on
the chain. The fact that the variables $r_{\Lm}$, $r_\Rm$ obey
Markovian stochastic differential equations is a consequence
of our choice of coupling functions $\rho_\L$, $\rho_\R$. In fact,
these functions are chosen in such a way that the random forces
exerted by the heat baths are not white noises but have covariances
which are (sums of) exponentials. Together
with the fluctuation theorem relating these random forces with the
dissipative forces, one obtains Markovian differential equations on
the phase space consisting of the physical variables $p$, $q$, augmented
by the auxiliary variables $r_{\Lm}$, $r_\Rm$.

\REMARK(n2) If the temperatures of the two heat baths are the same,
\ie, if $T_\L=T_\R$, the stationary state of the Markov process which
solves \equ(eqmo3) can be written explicitly. It is given by the
generalized Gibbs state
$$
\mu(\ddpqr)\,=\,\mu_{T_\L,T_\L}(\ddpqr)
\,=\, Z^{-1} e^{- G^{(0)}(r,q,p)/T_\L } \dpqr~,
\EQ(gibbs)
$$
where the ``energy'' $G^{(0)}$ is given by
$$
G^{(0)}(r,q,p)\,=\, H_\S(q,p) + \sum_{m=1}^{M}
\bigl( \frac{r_{\Lm}^2}{2\lambda_{\Lm}^2}
+ \frac{r_{\Rm}^2}{2\lambda_{\Rm}^2} -q_1\cdot r_{\Lm} - q_n\cdot r_{\Rm}
\bigr)~.
\EQ(gdef0)
$$
The marginal of this measure on the physical phase space is given
by
$$
\nu(\ddpq)\,=\, \int  \mu(\ddpqr) \,=\, {1\over Z'}
e^{- H_{\rm eff}/T_\L } \dpq ~,
$$
where the effective Hamiltonian $H_{\rm eff}$ is given by
$$
\eqalign{
H_{\rm eff}(q,p)\,&=\, H_{\S}(q,p)- \HALF q_1^2\smm\lambda_\Lm^2 -
\HALF q_n^2 \smm \lambda _\Rm^2\,\equiv\, \HALF p^2 +V_\eff(q)~.
}
\EQ(effham)
$$
It can be seen from \equ(effham) that the coupling between the chain
and the heat baths induces a renormalization of the potential $V(q)$.
In particular, because of Condition {\bf H1}, if the coupling
constants $\lambda_{im}$ are too large, $V_{\eff}(q)$ is not confining
any more, and the measure $\mu$ is not a probability measure, but only
$\sigma$-finite. In the sequel we require the following:
\item{\bf H3)} The coupling constants $\lambda_{im}$, $i\in\{\L,\R\}$,
$m \in \{1,\dots, M\}$ are such that
$$
\lim_{|q|\rightarrow \infty} V_{\rm eff}(q) \,=\, \infty ~.
$$

\SECTION Uniqueness of the Stationary State

In [EPR] we proved, under
the conditions {\bf H1}--{\bf H3}, the existence of an invariant
measure for any temperature $T_\L$, $T_\R$, but the uniqueness was
shown only for
small temperature differences. In this section, we extend the uniqueness
result
to {\em arbitrary} temperature
differences.

The uniqueness will follow from a dynamical argument: we will show
that the Markov process is transitive. This is done using a
(well-known) relationship
between stochastic differential equations and control theory (see
\eg, [Ku] and references therein).

We explain the method for a
general stochastic differential equation of the form
$$
\d x(t)\,=\, b(x(t))\dt  + \sigma dw(t)~,
\EQ(steq)
$$
where $x\in \real^k$, $b(x)$ is a $\CC^{\infty}$ vector field, $w(t)$
is a standard $\ell$-dimensional Brownian motion, and $\sigma$ is a
$k\times \ell$ matrix. We assume that the vector field $b(x)$ is such
that \equ(steq) has a unique solution for all $t>0$.
One then replaces $dw(t)$ in \equ(steq) by $u(t)\dt$. The function
$u(t)=(u_1(t), \dots, 
u_\ell(t))$ is called a control. One obtains the
system of ordinary differential equations
$$
{\dot x}\,=\, b(x(t)) + \sigma u(t) ~.
\EQ(cteq)
$$

The correspondence between the two systems is established by the
following result of Stroock and Varadhan [SV].
We fix an arbitrary time $\tau >0$.
Let $\UU$
denote the set of piecewise constant functions $u:[0,\tau
]\to \real^\ell$.
Let $\WW$
be the set of all continuous functions $\varphi$ from $[0,\tau]$ to $\real^k$
equipped with the uniform topology and let
$\WW_x=\{\varphi\in\WW~:~\varphi(0)=x\}$.
We denote $\xi_x$ the
diffusion process defined by \equ(steq) with initial condition
$\xi_x(0)=x$.
Then the path $\xi_x$ belongs almost surely to $\WW_x$.
The support of
this diffusion process $\xi_x$ on $[0,\tau ]$ is, by definition, the smallest
closed subset $\SS_x$ of $\WW_x$ such that
$$
{\bf P} [ \xi_x \in \SS_x ]\,=\, 1 ~,
\EQ(bigP)
$$
where {\bf P} is the probability induced by the Brownian motion
$w$.
We denote by $\varphi^u_x:[0,\tau ]\to \real^k$ the solution 
of the
differential equations \equ(cteq) with control $u$ and initial
condition $x$.
We next consider the notion of accessibility.
Let $x$ and
$y$ be two points in $\real^k$. The point $y$ is called {\it accessible}
from $x$ at time $\tau $ ($\tau >0$) if there is a control $u$ such that
$\varphi^u_x(\tau )\,=\,y$. The set of all points which are accessible from
$x$ at time $\tau $  is denoted $Y_\tau (x)$.

\CLAIM Theorem(strovarku) [SV]. One has
$$
\SS_x\,=\, { \overline { \{ \varphi_x ^u \, : \, u \in \UU ~\} } }~,
\EQ(allxt)
$$
for all $x\in\real^k$,
and
$$
 \supp
P(\tau ,x, \cdot)\,=\,{\overline{Y_\tau (x)}}~,
$$  
for all $x\in\real^k$ and $\tau >0$, where $P(\tau ,x,\d y)$ denotes
the
transition probability of the process $\xi_x$. 

\LIKEREMARK{Remark}The first statement is explicit in [SV] and the
second is a straightforward consequence of the first.

The main technical result of this section is the following
\CLAIM Theorem(chc) The control system associated with the
stochastic differential equation \equ(eqmo3) is strongly completely
controllable, \ie, $\overline{Y_\tau (x)}=\real^{2d(n+M)}$, for all
$x=(q,p,r_\L,r_\R)$ and all $\tau >0$.

\REMARK(now)We will combine this result with \clm(strovarku) and
hypoellipticity
to show that the invariant measure has a smooth, strictly positive density.

\REMARK(s1a)It should be noted that strong complete controllability
(SCC) can not
be deduced from
H\"ormander's hypoellipticity condition alone. (See \eg, [IK] for
examples of hypoelliptic diffusions with two invariant measures).
Therefore, \clm(chc) contains additional information.
Various sufficient conditions for SCC have been
expressed in terms of differential geometry in [Ku],
but these are {\em not} applicable to Eqs.\equ(eqmo3).

\LIKEREMARK{Proof of \clm(chc)}We will show the strong complete
controllability of the control problem associated with \equ(eqmo3)
by an explicit approach using the requirement of effective coupling (condition
{\bf H2}) of the chain. 

We reconsider the stochastic differential equation \equ(eqmo3).
Following the procedure described above we replace the Brownian
motions $w_{im}(t)$ by controls $u_{im}(t)$ in \equ(eqmo3),
and rewrite the system thus obtained as a system of second (and first)
order
equations. This leads to:
$$
\eqalignno{
\dot{r}_\Lm  \,&=\, -\gamma_\Lm  r_\Lm + \lambda_\Lm^2 \gamma_\Lm
q_1 +  u_\Lm ~, \quad m=1,\dots,M ~, \NR(f0)
\ddot q_1\,&=\,  -\nabla_{q_1}V(q) +\smm r_\Lm~,\NR(f1)
\ddot q_j\,&=\,  -\nabla_{q_j}V(q)~, \qquad \qquad j=2,\dots, n-1 ~, \NR(fj)
\ddot q_n\,&=\,  -\nabla_{q_n}V(q) +\smm  r_\Rm  ~,\NR(fn)
\dot{r}_\Rm  \,&=\, -\gamma_\Rm  r_\Rm  +
\lambda_\Rm^2\gamma_\Rm q_n + u_\Rm~, \quad m=1,\dots,M ~.  \NR(fn1)
}
$$
Here, we have absorbed the constants in front of the Brownian motion
in \equ(eqmo3) into the controls $u_{im}(t)$.   
We will only consider controls $u$ of class $\CC^\infty$. Any such function
can be uniformly approximated, on any compact interval, by a piecewise
constant function. Since a simple Gronwall estimate of Eq.\equ(cteq) shows
that
$$       
\sup_{t\in[0,\tau]}
|\varphi_x^u(t)-\varphi_x^v(t)|\,\le\, C(\tau)
                                    \sup_{t\in[0,\tau]}|u(t)-v(t)|
$$
holds with a constant $C(\tau)$ depending on the model, but not on $u$ and $v$,
we conclude that
$$   
\bigl \{\varphi_x^u(\tau)  : u\in \CC^\infty(\real)\bigr\} \,\subset\, \overline{ Y_\tau(x)}~.
$$
The proof of \clm(chc)
will now be done in two parts:
\LIKEREMARK{Part 1: Boundary control of the chain}We start by
considering the auxiliary problem of controlling a chain of $  n+2 $  
oscillators by the motion of the two ends of the chain.

The differential equation
reads
$$
\ddot{q}_{j}\,=\,f_{j}(q_{j-1},q_{j},q_{j+1})~,\quad j=1,\dots ,n ~,
\EQ(Dequ)
$$
 where $  q\equiv (q_{1},\dots ,q_{n}) $  is the dynamical variable, whereas
$  q_{0}\equiv u_{L} $  and $  q_{n+1}\equiv u_{R} $  are the control
variables. The smooth functions $  f_{j} $  are given by
$$
f_{j}(x,y,z)\equiv -(\nabla U_{j}^{(1)})(y)-(\nabla
U_{j}^{(2)})(y-z)+(\nabla U_{j-1}^{(2)})(x-y)~.
$$
Here, we define $  U_{0}^{(2)}(x)=U_{n}^{(2)}(x)=x^{2}/2 $.

Note that by Condition {\bf H2}, the functions $  \nabla U_{j}^{(2)} $  are diffeomorphisms.
It follows that the equation $  w=f_{j}(x,y,z) $  can be solved for $  z $ :
There exist smooth functions $  g_{j} $  such that $  w=f_{j}\bigl
(x,y,g_{j}(x,y,w)\bigr ) $ 
for all $  x,y,w\in \real ^{d} $. Consequently the differential equation \equ(Dequ)
is equivalent to the equation
$$
q_{j+1}\,=\,g_{j}(q_{j-1},q_{j},\ddot{q}_{j})~,\quad j=1,\dots ,n ~.
\EQ(Deltaequ)
$$
Obviously, for given $  q_{0} $  and $  q_{1} $, this equation can be solved
inductively, and in a unique way. To express this solution, let us introduce
some notation. For a smooth function $  \varphi  $  and an integer $  \alpha  $,
we shall denote the collection of the first $  \alpha  $  derivatives of $  \varphi  $
by $  \varphi ^{[\alpha ]}\equiv (\varphi ,\dot{\varphi },\dots
,\varphi ^{(\alpha )}) $. 
We also set $  p_{j}\equiv \dot{q}_{j} $  for $  j=1,\dots ,n $. For
$  q_{0}=\xi  $  
and $  q_{1}=\eta  $, the inductive solution of Eq.\equ(Deltaequ) reads
$$
\matrix{
u_{L} &\,=\,&  \xi  &\,\equiv\,&   G_{0}(\xi ^{[0]})~,\cr
q_{1} &\,=\,&  \eta  &\,\equiv\,&    G_{1}(\eta ^{[0]})~,\cr
p_{1} &\,=\,&  \dot{q}_{1} &\,\equiv\,&    G_{2}(\eta ^{[1]})~,\cr
q_{2} &\,=\,&  g_{1}(q_{0},q_{1},\dot{p}_{1}) &\,\equiv\,&   G_{3}(\xi
^{[0]},\eta ^{[2]})~,\cr 
p_{2} &\,=\,&  \dot{q}_{2} &\,\equiv\,&    G_{4}(\xi ^{[1]},\eta ^{[3]})~,\cr
q_{3} &\,=\,&  g_{2}(q_{1},q_{2},\dot{p}_{2}) &\,\equiv\,&    G_{5}(\xi
^{[2]},\eta ^{[4]})~,\cr 
 &\vdots &   & \vdots &   \cr
u_{R}  &\,=\,&  g_{n}(q_{n-1},q_{n},\dot{p}_{n}) &\,\equiv\,&    G_{2n+1}(\xi
^{[2n-2]},\eta ^{[2n]})~. \cr
}
$$
We can organize the $  2n+2 $  maps $  G_{J} $  into a map $  G\colon \real ^{4nd}\rightarrow \real ^{4nd} $ 
in the following way: Denote by $  (a,b) $  a point of $  \real ^{4nd} $, with
$  a\equiv (a_{0},\dots ,a_{2n-2})\in \real ^{(2n-1)d} $  and $  b\equiv (b_{0},\dots ,b_{2n})\in \real ^{(2n+1)d} $.
With $  a^{[\alpha ]}\equiv (a_{0},\dots ,a_{\alpha }) $  and $  b^{[\alpha ]}\equiv (b_{0},\dots ,b_{\alpha }) $,
define $  G(a,b)\equiv \bigl (a,\hat{G}(a,b)\bigr ) $  where 
$$
\hat{G}(a,b)\equiv
\bigl (G_{1}(b^{[0]}),G_{2}(b^{[1]}),G_{3}(a^{[0]},b^{[2]}),\dots
,G_{2n+1}(a^{[2n-2]},b^{[2n]})\bigr ) ~.
$$
We have proved that, if $  (u_{L},q_{1},\dots ,q_{n},u_{R}) $  is a solution
of Eq.\equ(Dequ) on the time interval $  I\subset \real  $, then
$$
(u_{L}^{[2n-2]},q_{1},\dot{q}_{1},\dots
,q_{n},\dot{q}_{n},u_{R})\,=\,G(u_{L}^{[2n-2]},q_{1}^{[2n]}) 
\EQ(MagicFormula)
$$
holds on $  I $. A simple consequence of this fact is that $  G $  is
a bijection. 
Indeed, repeated differentiation of Eq.\equ(Dequ) gives 
$$
\matrix{
q_{1}^{(2)}&\,=\,& f_{1}(u_{L},q_{1},q_{2}) &\,=\,&
F_{2}(u_{L}^{[0]},q_{1},q_{2})~, \cr 
q^{(3)}_{1}&\,=\,& \partial _{t}q_{1}^{(2)}
&\,=\,&F_{3}(u_{L}^{[1]},q_{1},\dot{q}_{1},q_{2},\dot{q}_{2}) ~, \cr 
  &\vdots &   & \vdots   \cr
q^{(2\alpha )}_{1}&\,=\,& \partial _{t}q^{(2\alpha -1)}_{1} &\,=\,&
F_{2\alpha }(u_{L}^{[2\alpha -2]},q_{1},\dot{q}_{1},\dots ,q_{\alpha
+1}) ~, \cr 
q^{(2\alpha +1)}_{1}&\,=\,& \partial _{t}q_{1}^{(2\alpha )} &\,=\,&
F_{2\alpha +1}(u_{L}^{[2\alpha -1]},q_{1},\dot{q}_{1},\dots,
q_{\alpha +1},\dot{q}_{\alpha +1}) ~, \cr  
  &\vdots &   & \vdots   \cr
q^{(2n)}_{1}&\,=\,& \partial _{t}q^{(2n-1)}_{1} &\,=\,&
F_{2n}(u^{[2n-2]}_{L},q_{1},\dot{q}_{1},\dots ,u_{R}) ~, \cr
}
$$
and thus we find another functional relation $
q_{1}^{[2n]}=\hat{F}(u_{L}^{[2n-2]},q_{1},\dot{q}_{1},\dots
,q_{n},\dot{q}_{n},u_{R}) $  
for its solutions. It immediately follows that the map $  F\colon
(a,b)\mapsto \bigl (a,\hat{F}(a,b)\bigr ) $  
satisfies 
$$
\eqalign{
F\circ G(u_{L}^{[2n-2]},q^{[2n]}_{1}) \,&=\,
(u_{L}^{[2n-2]},q^{[2n]}_{1})~, \cr
G\circ F(u^{[2n-2]}_{L},q_{1},\dot{q}_{1},\dots, u_{R}) \,&=\, 
(u^{[2n-2]}_{L},q_{1},\dot{q}_{1},\dots ,u_{R}) ~, 
}
$$
on every solution of Eq.\equ(Dequ). Since $  u_{L}^{[2n-2]}(0) $  and
either $  q_{1}^{[2n]}(0) $  or $  \bigl
(q(0),\dot{q}(0),u_{L}(0)\bigr ) $  can be prescribed
arbitrarily, we conclude that $  F=G^{-1} $.

To solve our control problem it suffices to remark that the set of solutions
$  (u_{L},q,u_{R}) $  satisfying $
\bigl (u^{[2n-2]}_{L}(t_{0}),q_{1}(t_{0}),\dot{q}_{1}(t_{0}),\dots
,u_{R}(t_{0})\bigr )=(a,b) $  
is identical with the set of solutions satisfying $
\bigl (u_{L}^{[2n-2]}(t_{0}),q^{[2n]}_{1}(t_{0})\bigr )=F(a,b) $. 
Since for $  \tau >0 $ and arbitrary $  (a,b),(a',b')\in \real ^{4nd} $  one
can find functions $  u_{L} $  and $  q_{1} $  for which 
$$
\eqalign{
\bigl (u_{L}^{[2n-2]}(0),q^{[2n]}_{1}(0)\bigr )\,&=\,F(a,b)~,\cr
 \bigl (u_{L}^{[2n-2]}(\tau ),q^{[2n]}_{1}(\tau )\bigr
 )\,&=\,F(a',b')~,\cr
}
$$
we see that the
system \equ(Dequ) 
is strongly controllable.
\LIKEREMARK{Part 2: Completion of the proof of \clm(chc)}We reduce the
problem of Eqs.\equ(f0)--\equ(fn1) to the case dealt with in Part 1,
by introducing the auxiliary variables $q_0$ and $q_{n+1}$.
Recalling the
definition $  U_{0}^{(2)}(x)=U_{n}^{(2)}(x)=x^{2}/2 $, we can rewrite
the control
problem associated with our stochastic differential equation as
$$
\eqalign{
\ddot{q}_{j} \,&=\,  f_{j}(q_{j-1},q_{j},q_{j+1})~,\quad j=1,\dots ,n ~, \cr
\sum ^{M}_{m=1}r_{L,m} \,&=\,  q_{1}-q_{0} ~,  \cr
\sum ^{M}_{m=1}r_{R,m} \,&=\,  q_{n+1}-q_{n} ~,    \cr
u_{L,m} \,&=\,  \dot{r}_{L,m}+\gamma _{L,m}r_{L,m}-\lambda
_{L,m}^{2}\gamma _{L,m}q_{1}~,\quad m=1,\dots ,M ~,  \cr
u_{R,m} \,&=\,  \dot{r}_{R,m}+\gamma _{R,m}r_{R,m}-\lambda
^{2}_{R,m}\gamma _{R,m}q_{n}~,\quad m=1,\dots ,M ~,  \cr 
}
\EQ(ControlPb)
$$
with the boundary conditions
$$
\eqalign{
\bigl (r_{L}(0),q_{1}(0),\dot{q}_{1}(0),\dots
,q_{n}(0),\dot{q}_{n}(0),r_{R}(0)\bigr ) \,&=\,  x ~, \cr
\bigl (r_{L}(\tau ),q_{1}(\tau ),\dot{q}_{1}(\tau ),\dots ,q_{n}(\tau
),\dot{q}_{n}(\tau ),r_{R}(\tau )\bigr ) \,&=\,  y ~.  \cr
}
$$
The equation $\sum ^{M}_{m=1}r_{L,m}=q_{1}-q_{0}$ serves to compensate
the term $q_1-q_0$ produced when differentiating $f_1$ in \equ(ControlPb).
Given the boundary data for $  r_{L} $, $  q_{1} $, $  q_{n} $  and $  r_{R} $ 
we obtain boundary values for $  q_{0} $  and $  q_{n+1} $. We can thus control
the first equation by our previous result. This gives us $
q_{0},\dots ,q_{n+1} $. 
Selecting arbitrary functions $  r_{L,2},\dots ,r_{L,M} $  and $
r_{R,2},\dots ,r_{R,M} $  
satisfying the corresponding boundary data, we define
$$
\eqalign{
r_{L,1} & \equiv  q_{1}-q_{0}-\sum ^{M}_{m=2}r_{L,m} ~,    \cr
r_{R,1} & \equiv  q_{n+1}-q_{n}-\sum ^{M}_{m=2}r_{R,m} ~.  \cr
}
$$
These two functions will also satisfy the boundary conditions. Finally we use
the last two sets of equations to determine the control variables $  u_{L,m} $ 
and $  u_{R,m} $.
This concludes the proof of \clm(chc).

\REMARK(s2)It is obvious from the proof of \clm(chc) that this theorem is
valid under much weaker conditions than those given in {\bf H1}.
It is enough to require that the stochastic differential
equation \equ(eqmo3) has a unique solution for all $t>0$. In particular we
do not need to restrict ourselves to potentials which are ``quadratic
at infinity'' as required in the proof of the existence of the
invariant measure.

The main result of this section is:
\CLAIM Theorem(unique++) If Conditions {\bf H1}--{\bf H3} are
satisfied, the Markov process which solves \equ(eqmo3) has a unique
invariant measure $\mu=\mu_T$. The measure $\mu$ has a $\CC^{\infty}$
density $\rho(r,q,p)$. This density is an exponentially
decaying, strictly positive function of $r$, $q$, and $p$. The
invariant measure is ergodic and mixing.

\REMARK(Decay)In fact, combining this result with information from
[EPR], one can show that
$$
\rho(r,q,p)\,=\,f(r,q,p)\exp\bigl (- G^{(0)}(r,q,p)/T^*\bigr )~,
$$
where $G^{(0)}$ was defined in
\equ(gdef0),
and $T^*=\max(T_\L,T_\R)$. The function $f$ is in the Schwartz space
$\SS$
when $T_\L\ne T_\R$, (and is a constant otherwise).

\PROOF The proof is a combination of \clm(strovarku) and \clm(chc)
with results in
[EPR]. The existence of the invariant measure $\mu$ is proven in [EPR,
Theorem 2.1].
Furthermore, using hypoellipticity,
we showed that the density $\rho$ of $\mu$ is $\CC^\infty
$.
Also, the transition probabilities $P(t,x,\d y)$ have
a smooth density, $p$, defined by $P(t,x,\d y) = p(t,x,y)\d y$ with
$p(t,x,y) \in
\CC^{\infty} ((0,\infty), \real^{2d(n+M)}, \real^{2d(n+M)})$.

We next show that the support of $\mu$ is all of the extended phase space
$X\equiv \real^{2d(n+M)}$. In \clm(chc), we
have seen that \equ(eqmo3) is strongly completely controllable.
By
\clm(strovarku), we conclude that
the support of $P(\tau,x,\cdot)$ is the whole phase
space for every $x\in X$ and all $\tau >0$.
Therefore, we have, for all $t>0$, all $x$, and all open sets $Y$,
the inequality
$$
P(t,x,Y)\,>\,0~.
$$
Since $\mu(Y)=\int \mu(\d x) P(t,x,Y) $ (because $\mu$ is invariant),
we conclude that $\supp \mu = X$ and thus
the density $\rho$ is Lebesgue almost everywhere positive.

We next show that $\rho(x) > 0$, for all $x$.
by assuming the contrary and deriving a contradiction.
Assume that there is a $y$ for which $\rho(y) = 0$.
By the invariance of the
measure we have, for any $t>0$,
$$
0\,=\,\rho(y)\,=\, \int \d x\,\rho(x)\, p(t,x,y)~.
$$
This implies
$p(t,x,y)\,=\,0$ for Lebesgue almost all $x$.
Since the transition kernel $p$ is smooth, we conclude that the
function
$p(t,\cdot,y)$ is identically zero every $t>0$.
On the other hand, since $p$ is the kernel of a strongly continuous
semigroup, we also have
$p(t,x,y) \rightarrow
\delta(x-y)$ as $t\rightarrow 0$.
This is a contradiction, and we have shown
$\rho(y)>0$, for all $y\in X$.

We next show uniqueness. We have just shown that every invariant
measure must have a smooth, strictly positive density.
Since every ergodic component is mutually singular to any other, the
invariant measure is unique (and ergodic).
The property of mixing of the invariant measure has been deduced from
uniqueness in the proof of
[EPR, Theorem 3.9] . This concludes the proof of \clm(unique++).

\REMARK(analyticity) We proved in [EPR, Lemma 3.7] that the
density $\rho=\rho_T$ is a real analytic function
of $\zeta=(T_\L-T_\R)/(T_\L+T_\R)$. In particular, this yields the
standard perturbative results near equilibrium ($\zeta=0$).

\SECTION Time-Reversal, Energy Flux, and Entropy Production

In this section, we ask questions which are intimately related to the
Hamiltonian nature of our model. After introducing appropriate
notation, we introduce time reversal, and draw some consequences. In
particular, we are able to show that the system exhibits non-zero mean
energy flux as soon as $T_\L\ne T_\R$, and we relate the flux to the
entropy production.

\SUBSECTION Notation

It will be useful to streamline the notation.
It is convenient to introduce first $\rho_\Lm =(\lambda_\Lm
\gamma_\Lm^{1/2})^{-1}r_\Lm$ and similarly for the $r_\Rm$. Then the
equations of motion are
$$\eqalign{
\d q_j (t) \,&=\, p_j (t)\dt~, \qquad\qquad\qquad j =1,\dots,n ~,\cr
\d p_1 (t) \,&=\, -\nabla_{q_1}V(q(t))\dt  +\smm \lambda_\Lm
\gamma_\Lm^{1/2}\rho_\Lm(t)\dt~,\cr
\d p_j (t) \,&=\, -\nabla_{q_j }V(q(t))\dt ~,\qquad j =2,\dots,n-1 {}~,\cr
\d p_n (t) \,&=\, -\nabla_{q_n}V(q(t))\dt  +\smm  \lambda_\Rm
\gamma_\Rm^{1/2}\rho_\Rm (t)\dt ~,\cr
\d \rho_\Lm (t) \,&=\, -\gamma_\Lm  \rho_\Lm(t)\dt + \lambda_\Lm
\gamma_\Lm^{1/2} q_1(t) \dt
                 - \sqrt{2}T_\L^{1/2}\, \d w_\Lm(t)  ~,\cr
\d \rho_\Rm (t) \,&=\, -\gamma_\Rm  \rho_\Rm (t)\dt + \lambda_\Rm
\gamma_\Rm^{1/2} q_n(t) \dt
                 - \sqrt{2}T_\R^{1/2}\, \d w_\Rm(t)~,  \cr
&\quad\quad\quad\quad m=1,\dots,M~.
}\EQ(eqmo3a)
$$

We can write this system in vector notation:
We write the Hamiltonian of the chain (the small system) as
$$
H_\S(q,p)\,=\,{p^2\over 2} + V(q)~,
$$
with $q,p\in \real^{nd}$. The two reservoirs, $\L$ and $\R$, are
described by the variables $\rho=(\rho_\L,\rho_\R)\in \real^{Md}
\oplus\real^{Md}$.
The ``energy'' of the complete system, \ie, chain and
reservoirs, is then given by $G^{(0)}(r,q,p)=G^{(1)}(\rho,q,p)$, where now
$$
G^{(1)}(\rho,q,p)\,=\,H_\S(q,p) + \HALF \rho\cdot \Gamma \rho- q\cdot \Lambda
\Gamma^{1/2} \rho~.
$$
Here, $\Gamma=\Gamma_\L \oplus\Gamma_\R$, where $\Gamma_i$ is the
diagonal ($M\times M$) matrix
${\rm diag}(\gamma_{i,1},\dots,\gamma_{i,M})$, with $i\in\{\L,\R\}$.
Note that by assumption, the $\gamma$'s are all strictly positive.
We also define $\Lambda$ as the ($2Md\times nd$) matrix given by
$$
q\cdot \Lambda \rho\,=\,
q_1\cdot \Lambda_\L \rho_\L + q_n \cdot \Lambda_\R \rho_\R
\,=\,
q_1\smm \lambda _\Lm\rho_\Lm
+q_n \smm \lambda _\Rm \rho_\Rm~.
$$
With these notations, the equations of motion can be written as:
$$
\eqalign{
\d q\,&=\,\nabla_p G^{(1)} \,\dt 	\,=\,  p\, \d t~,\cr
\d p\,&=\,-\nabla_q G^{(1)}\, \dt \,=\, -\bigl (\nabla_q V(q) - \Lambda \Gamma^{1/2}
\rho \bigr )\dt~,\cr
\d\rho\,&=\, -\nabla_\rho G^{(1)}\,\dt - (2T^{1/2}) \d w \,=\,
-\bigl (\Gamma\rho -\Gamma^{1/2}\Lambda^\T q\bigr )\dt  - (2T^{1/2}) \d w~. \cr
}
$$
Here, $w=w_\L\oplus
w_\R=(w_{\L,1},\dots,w_{\L,M},w_{\R,1},\dots,w_{\R,M})$ is a
$2Md$-dimensional standard Brownian motion, and $T$ is the
($2M\times2M$) diagonal temperature matrix
$$
T\,=\,{\rm diag}(T_\L,\dots,T_\L,T_\R,\dots,T_\R)~.
$$
It is useful to introduce the (final!) change of variables
$s=\rho - F^{\T} q$, where $F= \Lambda \Gamma^{-1/2}$. In terms of
these variables, one can introduce the effective potential
$$
V_\eff(q)\,=\,V(q) -\HALF q\cdot \Lambda \Lambda ^\T q~,
\EQ(Veff)
$$
and the ``energy'' is now $G(s,q,p)=G^{(1)}(\rho,q,p)$ with
$$
G(s,q,p)\,=\, \HALF p^2 +V_\eff+ \HALF s\cdot \Gamma s~.
\EQ(gdef)
$$
Finally, with the adjoint change in the derivatives $\nabla_q \to
\nabla_ q - F\nabla _s$, the equations of motion become
$$
\eqalign{
\d q\,&=\,\nabla_p G \dt 	\,=\,  p\, \d t~,\cr
\d p\,&=\,-(\nabla_q- F\nabla_s) G \dt \,=\, -\bigl (\nabla_q
V_\eff(q) - F \Gamma s
\bigr )\dt~,\cr
\d s\,&=\, -(\nabla_s +F^\T\nabla_p) G\,\dt - (2T^{1/2}) \d w \,=\,
-\bigl (\Gamma s+ F^\T p\bigr )\dt  - (2T^{1/2}) \d w~. \cr
}
\EQ(final)
$$

\LIKEREMARK{Notation}In the sequel, we shall write $G_p$ for $\nabla_p
G$ and $G_q$ for $\nabla_q G$ (these are vectors with $nd$
components),
and
$G_s$ for $\nabla_s G$ (this is a vector with $2Md$ components).

The generator $L$ of the
diffusion process takes, in the variables $s$, $q$, $p$, the form
$$
L\,=\,\nabla_s\cdot T\nabla_s - G_s \cdot \nabla_s + \bigl (
G_p\cdot \nabla_q-G_q\cdot\nabla_p
\bigr )
+\bigl ((FG_s) \cdot \nabla_p-G_p\cdot F\nabla_s\bigr )~.
\EQ(ldef)
$$
If $f$ is a function on the phase space $X$, we let
$$
S^t f(x)\,=\,\bigl (e^{Lt} f\bigr )(x)\,=\,
\int f\bigl (\xi_x(t)\bigr ) d{\bf P}(w)~.
$$
The associated Fokker-Planck operator $L^\T$ is the adjoint of $L$ in
the space $\L^2(\real^{d(2M+2n)},\d x)$, \ie,
$$
L^\T\,=\,\nabla_s\cdot T\nabla_s +\nabla_s\cdot G_s  - \bigl (
G_p\cdot \nabla_q-G_q\cdot\nabla_p
\bigr )
-\bigl ((FG_s) \cdot \nabla_p-G_p\cdot F\nabla_s\bigr )~.
\EQ(ltdef)
$$
\REMARK(r4)The density $\rho$ of the invariant measure is the (unique)
normalized
solution of
the equations
$$
L^\T \rho\,=\,0~.
$$

\SUBSECTION The Entropy Production $\sigma$

We now establish the relation between the
energy flux and the entropy production. Since we are dealing with a
Hamiltonian setup, the energy flux is naturally defined by the time
derivative of the mean evolution $S^t$ of the
effective energy, $H_\eff(q,p)= p^2/2+V_\eff(q)$.
Differentiating, we get from the equations of motion
$$
\partial_t  S^t H_\eff\,=\, S^t L  H_\eff~,
$$
$$
LH_\eff\,=\,  p\cdot (-\nabla_q V_\eff +  F\Gamma s)
+\nabla_q V_\eff\cdot p \,=\,p\cdot F\Gamma s~.
$$
We define the total flux by $\Phi=p\cdot F\Gamma s$, and inspection of
the definition of $F$ and $\Gamma $ leads to the identification of the
flux at the left and right ends of the chain:
$$
\Phi\,=\,\Phi_\L+\Phi_\R~,
$$
with
$$
\eqalign{
\Phi_\L\,&=\,p_1\cdot \Lambda _\L \Gamma _\L^{1/2} s_\L~,\cr
\Phi_\R\,&=\,p_n\cdot \Lambda _\R \Gamma _\R^{1/2} s_\R~.\cr
}
$$
Note that $\Lambda_\L\Gamma_\L^{1/2}s_\L$ is the net force exerted by
the
left bath on the chain. Therefore,
$\Phi_\L=p_1\cdot\Lambda_\L\Gamma_\L^{1/2}\rho_\L-L\,q_1\cdot\Lambda_\L^2q_1/2$
is, up 
to a time-derivative which vanishes in the stationary state, the
total power dissipated by the left bath. A similar interpretation holds
for $\Phi_\R$.
Furthermore, observe that
$$
\langle \Phi \rangle _\mu\,=\,0~,
\EQ(phii)
$$
where, generally,
$$
\langle f\rangle_\mu\,\equiv\,\int \mu(\d x) f(x)~.
$$
The Equation \equ(phii) holds because
$\Phi=L H_\eff$ and $L^\T \mu=0$.

We next proceed to define the entropy production in the setting of our
model. Since we have been able to identify the energy flux on the ends
of the chain, we {\em define} the (thermodynamic) entropy production
$\sigma$ by
$$
\sigma\,=\, {\Phi_\L\over T_\L}+ {\Phi_\R\over T_\R}\,=\, p\cdot F
T^{-1}\Gamma s~.
\EQ(entropy)
$$
We refer to [CL] and references therein 
for a detailed discussion of the various types of
entropy production in non-equilibrium stationary states. 
In Subsection 4.4, we will explain, in the context of our model, the
relationship between the 
entropy production $\sigma$ and the Gibbs entropy.

\SUBSECTION Time-Reversal, Generalized Detailed Balance Condition, and
Negativity of the Entropy Production

\LIKEREMARK{Definition}We define the ``time-reversal'' map $J$ by
$\bigl (Jf\bigr )(s,q,p)=f(s,q,-p)$. This map is the
projection onto the space of the $s,q,p$ of the
time-reversal of the
Hamiltonian flow (on the full phase space of chain plus baths)
defined by the original problem \equ(eqmo1).

\LIKEREMARK{Notation}To obtain simple formulas for the entropy
production $\sigma$
we write the strictly positive density $\rho$ of the invariant
measure $\mu$ as
$$
\rho\,=\,Je^{-R}e^{-\varphi}~,
\EQ(phidef)
$$
where we have introduced the quantity
$$
R\,=\,R(s)\,=\,\HALF s\cdot \Gamma T^{-1}s~.
\EQ(rdef)
$$
Let $L^*$ denote the adjoint of $L$ in the space $\HH_\mu=\L^2(X,\d
\mu)$ associated with the
invariant measure $\mu$ with density $\rho$. In terms of the
adjoint $L^\T$ on $\L^2(X,\dspq)$, we have
$$
L^*\,=\,\rho^{-1} L^\T \rho~.
\EQ(lstar)
$$
Let $L_\lambda = L
+\lambda \sigma$, where $\lambda \in \real$. (This definition is
suggested by the paper [K], see below.) 
We have the following important symmetry property:
\CLAIM Theorem(LS) One has the operator identity
$$
J e^{-J\varphi} (L_\lambda )^* e^{J\varphi} J \,=\, L_{1-\lambda }~.
\EQ(LS)
$$
In particular,  one has
$$
J e^{-J\varphi} L^* e^{J\varphi} J -L \,=\,\sigma~.
\EQ(LS0)
$$

\REMARK(DBC)This relation may be viewed as a generalization to
non-equilibrium of the detailed balance condition (at equilibrium,
one has $JL^*J-L=0$).

\REMARK(LS1)Recently, a lot of interest has been generated in the wake
of papers by Gallavotti and Cohen, [GC1, GC2, G, and references therein], in
which intriguing 
relations for the fluctuations of the entropy production have been
found.
These papers dealt first with numerical experiments by [ECM], which
were then abstracted to the general context of dynamical systems. In
further work, these ideas have been successfully applied to
thermostatted systems modeling non-equilibrium problems.
In the papers [K] and  [LS] these ideas have been further extended
to non-equilibrium models described by stochastic dynamics. 
In the context of our model, the setup is as follows:
One considers the observable
$$
W(t)\,=\,\int_0^t \d \tau\, \sigma\bigl (\xi_x(\tau )\bigr ) ~.
$$
By ergodicity,
$\lim _{t\to\infty } t^{-1}W(t)=\langle \sigma \rangle_\mu$,
for $\mu$-almost all $x$. 
We are interested in the rate function $\hat e$
for the large deviations of
$W(t)/(t\langle \sigma \rangle_\mu)$,
and want to argue (heuristically) that it
satisfies
$$
\hat e(w) - \hat e(-w)\,=\, -w  \langle \sigma \rangle_\mu~.
\EQ(symm1)
$$
In particular this means that at equal temperatures, when $\langle
\sigma \rangle_\mu=0$, the fluctuations are symmetric around the mean
$0$, while at unequal temperatures, the odd part is linear in $w$ and
proportional to 
the mean entropy production.
This is the celebrated Gallavotti-Cohen fluctuation theorem.

The rate function $\hat e$ is
characterized by the relation
$$
\inf _{w\in I} \hat e(w) \,=\,
-\lim_{t\to\infty } {1\over t} \log {\bf P}\left ( {W(t)\over t
\langle \sigma \rangle_\mu}\in I\right  )~.
$$
Under suitable conditions it can be expressed as the Legendre
transform of the function
$$
e(\lambda )\,\equiv\, -\lim_{t\to\infty } t^{-1}\log \bigl \langle
e^{-\lambda W(t)}\bigr \rangle_\mu~.
$$
Formally, $-e(\lambda )$ can be represented as the maximal
eigenvalue of $L_\lambda $. Observing now 
the relation \equ(LS), one sees immediately that
$$
e(\lambda)\,=\,e(1-\lambda )~.
\EQ(symm)
$$ 
This in turn implies \equ(symm1).
A rigorous derivation of
the program outlined above lacks several technical ingredients, in
particular, more spectral information about $L_\lambda $ seems to be necessary.

The relation \equ(LS) has a number of other consequences which we enumerate
now, before going to the proof of \clm(LS). It allows to prove
that the entropy production is negative in our model and it yields
an interesting symmetry relation (see \clm(entropy)).

\CLAIM Proposition(relations) One has the following identities
(between functions):
$$
\eqalignno{
L\varphi\,&=\,\hphantom{-}\sigma + |T^{1/2} \nabla_s \hphantom {J}
\varphi|^2~,\NR(lphi) 
L^*J\varphi\,&=\,-\sigma - |T^{1/2} \nabla_s J
\varphi|^2~.\NR(lstarphi)
}
$$

Here, $|f|^2\equiv f\cdot f$.

\CLAIM Theorem(entropy) In the stationary state $\mu$ the entropy
production satisfies the identity:
$$
\langle  \sigma \rangle _\mu \,=\,
-\langle | T^{1/2} \nabla_s \varphi |^2 \rangle_\mu  \,=\,
-\langle | T^{1/2} \nabla_s J\varphi |^2 \rangle_\mu  \,\le\, 0 ~.
\EQ(ident1)
$$

\REMARK(positive)In Subsection 4.5, we will show that the heat flux is
non-zero for unequal temperatures by showing that the entropy
production in the stationary state satisfies:
$$
\langle  \sigma \rangle _\mu \,=\, 0 ~,
$$
if and only if\/ $T_\L = T_\R$.

The remainder of this subsection is devoted to the proofs of \clm(LS),
\clm(relations), and \clm(entropy).

\LIKEREMARK{Proof of \clm(LS)}We show the identity
$$
e^{R}JL^{\T}Je^{-R}\,=\,L+\sigma  ~.
\EQ(id1)
$$
Starting with the relation
$$
e^R\nabla_s e^{-R}\,=\,\nabla_s -(\nabla_s R)\,=\,\nabla_s
-T^{-1}G_s~,
$$
we get, using the definition of $L^\T$,
$$
\eqalign{
e^R J L^\T J e^{-R}\,&=\,
\nabla_s\cdot T\nabla_s -  G_s\cdot\nabla_s
\cr
& +\bigl (
G_p\cdot \nabla_q-G_q\cdot\nabla_p
\bigr )
+\bigl (F\Gamma s \cdot \nabla_p-p\cdot F\nabla_s\bigr )
\cr &+G_p\cdot FT^{-1}G_s~.\cr
}
$$
Note that the sum of all the terms except the last equals $L$, while the
last equals
$$
G_p\cdot FT^{-1}G_s\,=\,p\cdot F T^{-1}\Gamma s\,=\,
{p_1 F_\L \Gamma _\L s_\L\over T_\L}
+{p_n F_\R \Gamma _\R s_\R\over T_\R}\,=\, \sigma ~.
$$
We have thus shown \equ(id1).
Combining \equ(id1) with the expressions \equ(lstar) and \equ(phidef)
for $L^*$ and $\rho$ we 
obtain the identity:
$$
L+\sigma\,=\, e^R J L^\T J e^{-R}\,=\, e^R J e^{-R} e^{-J\varphi} L^*
e^{J\varphi} e^R J e^{-R}  \,=\, J e^{-J\varphi} L^* e^{J\varphi}J~,
\EQ(LS1)
$$
which is \equ(LS) for $\lambda =0$, \ie, Eq.\equ(LS0).
Observing now that $J\sigma J = -\sigma$, we obtain 
$$
J e^{-J\varphi} L_\lambda^* e^{J\varphi}J\,=\,L_{1-\lambda}~,
$$
and thus conclude the proof of \clm(LS).

\LIKEREMARK{Proof of \clm(relations)}From \equ(LS) we obtain the identity
$$
JL^*J\,=\,e^{\varphi}(L+\sigma)e^{-\varphi}~.
\EQ(uuu)
$$
A straightforward computation shows that, for any smooth function $f$,
we have the following operator identity 
$$
e^{f }L e^{-f }\,=\, L-
2(\nabla_sf )\cdot T\nabla_s -  (Lf)  + |T^{1/2}(\nabla_sf )|^2  ~.
\EQ(id2)
$$
Applying \equ(id2) with $f=\varphi$ we obtain from Eq.\equ(uuu) the
operator identity
$$
JL^*J\,=\, L - 2(\nabla_s\varphi )\cdot T\nabla_s -  (L\varphi)  +
|T^{1/2}(\nabla_s\varphi )|^2 + \sigma ~. 
\EQ(i3)
$$
Since $L^*=\rho^{-1}L^\T \rho$ and $L^\T\rho=0$ we have $L^*1=0$. 
Applying the operator identity \equ(i3) to the function $1$ and noting
that $J1=1$ we get
$$
0\,=\,JL^*J1\,=\,-L\varphi + |T^{1/2}(\nabla_s\varphi )|^2 + \sigma ~,
\EQ(i4)
$$
and this is the identity \equ(lphi). With this, \equ(i3) simplifies to
$$
JL^*J\,=\,L - 2(\nabla_s\varphi )\cdot T\nabla_s~.
\EQ(i5)
$$
Applying the operator identity \equ(i5) to the function $\varphi$, and
using \equ(i4) we get
$$
\eqalign{
JL^*J \varphi \,&=\, L\varphi - 2 |T^{1/2}(\nabla_s\varphi )|^2 \cr
              \,&=\, \sigma -  |T^{1/2}(\nabla_s\varphi )|^2~, \cr
}
$$
or, equivalently,
$$
L^*J \varphi\,=\, -\sigma - |T^{1/2}(\nabla_sJ\varphi )|^2 ~,
$$
which proves \equ(lstarphi).
With this we have concluded the proof of \clm(relations).

\LIKEREMARK{Proof of \clm(entropy)}\clm(entropy) is a simple consequence of
\clm(relations). From \equ(lphi), using the invariance of the measure
$\mu$, we get
$$
\eqalign{
\langle \sigma \rangle_\mu \,&=\, \langle L\varphi \rangle_\mu - 
\langle |T^{1/2}(\nabla_s\varphi )|^2 \rangle_\mu \cr
\,&=\, - \langle |T^{1/2}(\nabla_s\varphi )|^2 \rangle_\mu ~,\cr
}
$$
which yields the first equality in \equ(ident1). The second inequality
is obtained in the same way using \equ(lstarphi). We have  
$$
\eqalign{
\langle \sigma \rangle_\mu \,&=\, - \langle L^* J\varphi \rangle_\mu - 
\langle |T^{1/2}(\nabla_sJ\varphi )|^2 \rangle_\mu \cr
\,&=\, - \langle |T^{1/2}(\nabla_sJ\varphi )|^2 \rangle_\mu ~,\cr
}
\EQ(i23)
$$
where the last equality in \equ(i23) follows from the identity
$$
\langle L^* J\varphi \rangle_\mu \,=\, \int \, \d x  (L^\T \rho
J\varphi) \,=\, \int \d x \rho J\varphi (L1)\,=\,0~. 
$$
It is obvious from \equ(i23) that the entropy
production in the stationary state is a non-positive quantity and
this concludes the proof of \clm(entropy).

\LIKEREMARK{Other observables for the entropy production}The analysis
done for the entropy production $\sigma$ can be repeated for other
observables, (see also [LS] for a similar generalization).
A family of such observables can be obtained  by
replacing the conjugation operator $e^{J\varphi}J$ of \equ(LS) by any conjugation
operator of the form $e^f e^{J\varphi}J$, where $f=f(q,p)$ satisfies
$Jf=f$.\footnote{${}^1$}{The operators $e^f e^{J\varphi}J$ are all formally
selfadjoint on $\HH_\mu$.} 
Interesting examples are obtained when one considers the energy flux
between position 
$j$ and the $j+1$ on the chain, $j=1,\dots,n-1$.
To this end we write the
Hamiltonian $H_\eff$ as follows:
$$
H_\eff(q,p)\,=\,\sum_{i=1}^n H_i(q,p)~,
$$
where
$$
\eqalign{
H_1(q,p) \,&=\, {p_1^2\over 2} + U^{(1)}_1(q_1) - {1\over 2} q_1^2\sum_{m=1}^{M}  \lambda^2_\Lm
+  {1\over 2}U_1^{(2)}(q_1-q_2) ~,\cr 
H_i(q,p) \,&=\, {p_i^2\over 2} + U^{(1)}_i(q_i) + 
 {1\over 2}U_{i-1}^{(2)}(q_{i-1}-q_{i}) +   {1\over 2}U_i^{(2)}(q_{i}-q_{i+1})~, \quad i=2,
\dots, n-1  ~, \cr 
H_n(q,p) \,&=\, {p_n^2\over 2} + U^{(1)}_n(q_n) - {1\over 2} \sum_{m=1}^{M} \lambda^2_\Rm \,q_n^2
+  {1\over 2} U_{n-1}^{(2)}(q_{n-1}-q_n) ~.\cr 
}
$$
For any $j=1,\dots,n-1$, we choose $f=-S_j$, where
$$
S_j(q,p) \,=\, \frac{1}{T_\L} \sum_{i=1}^j H_i(q,p) + 
\frac{1}{T_\R} \sum_{i=j+1}^n H_i(q,p)~.
$$
We write now the invariant density $\rho$ as
$$
\rho\,=\, J e^{-R} e^{-S_j} e^{-\psi_j}~,
$$
\ie, $\psi_j=\varphi - S_j$.
Variants of computations done above show that we
have the operator identity, similar to \equ(LS1):
$$
e^{S_j} e^R J L^\T J e^{-R} e^{-S_j}\,=\, L + \sigma_j~,
$$
where $\sigma_j$ is given by
the relation
$$
\sigma_j\,=\, \sigma - LS_j~,
\EQ(ssj)
$$
since $S_j$ does not depend on the variable $s$.
Our choice of $S_j$ has been made in such a way that
$$
\sigma_j\,=\, \bigl( \frac{1}{T_\L}-\frac{1}{T_\R}\bigr)\,\, (p_j-p_{j+1})\cdot
\nabla U^{(2)}(q_j-q_{j+1})~,
$$
\ie, $\sigma_j$ is the energy flux between position $j$ and
$j+1$ on the chain multiplied by the difference between the
inverse temperatures of the heat baths.
Using next that $JS_j=S_j$ one
derives easily a relation corresponding to \equ(LS), 
namely,
$$
J e^{-J\psi_j} (L+\lambda \sigma_j )^* e^{J\psi_j} J \,=\, L+
(1-\lambda)\sigma_j ~.
\EQ(LSj)
$$

We have thus found $n-1$ ``entropy productions'' $\sigma_j$, which, under the
assumptions made for $\sigma$, satisfy a fluctuation theorem. Note
that these entropy productions are all different observables, but,
because of Eq.\equ(ssj),
the
expectations of $\sigma$ and $\sigma_j$ in the
stationary state $\mu$ coincide.

\SUBSECTION Relation with the Gibbs Entropy

We give now a second proof of the negativity of entropy production in
our model using the Gibbs entropy.

Let $\nu_0$ be a probability measure in the variables
$x=(s,q,p)$ and let $\nu_t$ denote the time evolution of $\nu_0$ given
by
$$
\nu_t(\d x)\,=\, \int \nu_0(\d x') P(t,x', x)~.
$$
Because of the properties of the transition probabilities
$P(t,x', x)$ proven in Sect.~3, we see that
$\nu_t$ is a probability measure (for any $t>0$)
with a smooth positive density
denoted $f_t$ in the sequel.
The time evolution of $f$ is then given by the equation
$$
\partial_t f_t \,=\,L^\T f_t~.
$$
We define the Gibbs entropy as
$$
S(f)\,=\,-\int \d x \,f(x) \log f(x)~,
$$
and we compute next the entropy change in time.
We get:
$$
\eqalign{
\partial_t S(f_t)\,&=\,-\int \d x\, (\partial_t f_t) (1+\log f_t)\cr
\,&=\, - \bigl ( L^\T f_t, (1+\log f_t ) \bigr )\cr
\,&=\, - \bigl ( f_t , L \log f_t \bigr )\cr
\,&=\, -\bigl (f_t, f_t^{-1} L f_t \bigr ) + ( f_t,
|T^{1/2}\nabla_s \log f_t|^2\bigr )\,\equiv\,X_1~.
}
$$
The last term is the (additional) contribution from the second order
derivative (in
$s$) appearing in $L$ when it acts on $\log f$.
We can transform $X_1$ further by writing it as
$$
\eqalign{
X_1\,&=\, -(L^\T 1, f_t) + \bigl ( f_t, | T^{1/2} \nabla_s \log f_t |^2\bigr
)~.
}
\EQ(e1)
$$
Since $L^{\T}1=\Tr \Gamma$ the first term in \equ(e1) is equal
to $-\Tr \Gamma$.
We use the definition \equ(rdef) of $R$ and the analog of
\equ(phidef) to define $\varphi _t$:
$$
J e^{-R} e^{-\varphi_t }\,=\,f_t ~.
\EQ(phidef3)
$$
Since $JRJ=R$, we see that $-\log f_t= R+J\varphi _t$.
Expanding the second term in \equ(e1)  we obtain
$$
\eqalign{
( f_t, | T^{1/2} \nabla_s \log f_t |^2\bigr)\,&=\,
( f_t, | T^{-1/2} \Gamma s |^2\bigr) +
( f_t, | T^{+1/2} \nabla_s J \varphi_t |^2\bigr) +
2 ( f_t, \Gamma s \cdot \nabla_s J \varphi_t) ~.
}
\EQ(e2)
$$
Since we have the relation
$$
\nabla_s f_t \,=\, -f_t \bigl( \nabla_s R + \nabla_s J \varphi_t
\bigr)~,
$$
we obtain
$$
f_t \nabla_s J \varphi_t \,=\, -f_t  \nabla_s R  - \nabla_s f_t~.
$$
Using this and integrating by parts,
we rewrite the third term in \equ(e2) as
$$
2 ( f_t, \Gamma s \cdot \nabla_s J \varphi_t) \,=\,
-2 \int \d x\,\Gamma s \cdot \bigl(f_t \nabla_s R  + \nabla_s f_t  \bigr)\,=\,
 - 2 (f_t, |T^{-1/2} \Gamma s|^2)+2 \Tr \Gamma~.
$$
Altogether we obtain
$$
\partial_t S(f_t) \,=\, \Tr\Gamma  - (f_t, |T^{-1/2} \Gamma s|^2) +
( f_t, | T^{1/2} \nabla_s J \varphi_t |^2\bigr)~.
$$
Using the identity
$$
L R \,=\, \Tr\Gamma  -  |T^{-1/2} \Gamma s|^2 -\sigma ~,
$$
we obtain finally,
$$
\partial_t S(f_t) \,=\, \int \d x\,f_t \sigma + \int  \d x\,f_t LR + \int \d x\, f_t | T^{1/2} \nabla_s J
\varphi_t |^2 ~.
\EQ(varentropy)
$$
In line with the ideas of [CL], we can write this last identity in the
form:
$$
\partial_t S(f_t)-\langle \sigma\rangle_{\nu_t} \,=\, \bigl \langle | T^{1/2} \nabla_s J
\varphi_t |^2 \bigr \rangle_{\nu_ t} + \partial_t \langle R\rangle_{\nu_t}~.
$$
This shows that the (rearrangement) entropy produced in addition to
the thermodynamic entropy $\sigma$ is a positive quantity, up to a
(time-) boundary term.
Also note that if $\nu_t=\mu$, \ie, if the system is in the stationary
state, then we get the identity
$$
0\,=\,\langle
\sigma\rangle_{\mu}
+\langle
|T^{1/2}J\nabla_s\varphi|^2\rangle_{\mu}~,
\EQ(ident2)
$$
which we already found in \clm(entropy).
\SUBSECTION Strict Positivity of the Heat Flux

We first show that the thermodynamic entropy production, as defined in 	
\equ(entropy), is negative in our model. As an immediate consequence we
will show that, in the stationary state, energy is flowing from the
hotter heat bath to the colder one.

\CLAIM Theorem(positive1) The entropy production $\sigma$ satisfies:
$$
\langle  \sigma \rangle _\mu \,=\, 0 ~,
$$
if and only if\/ $T_\L = T_\R$.

\PROOF Note that if $T_\L = T_\R$, then $\sigma= \left .\partial_t S^t
H_{\rm eff}/T_\L\right |_{t=0}$
and therefore $\langle \sigma \rangle_\mu =0$.  We will show that if
$T_\L\ne T_\R$, then $\langle \sigma \rangle_\mu\ne 0$.
We will proceed by assuming the converse, namely $\langle \sigma
\rangle_\mu= 0$, and produce a contradiction.
The assumption implies, by \equ(ident1) that $\langle | T^{1/2}
\nabla_s \varphi |^2 \rangle_\mu=0$. Since $\rho$ is positive, this
means that
$\nabla_s \varphi=0$, and therefore $\varphi $ does not
depend on the $s$ variables. From \equ(lphi) we obtain
$$
\eqalign{
0\,&=\, -L\varphi + |T^{1/2}\nabla_s \varphi |^2 + \sigma
 \,=\, -L\varphi + \sigma ~.
}
$$
Using the definition of $L$ and $\sigma$ and the fact that $\varphi$
does not depend
on $s$, we obtain the equation
$$
0\,=\, \left( p\cdot \nabla_q \varphi - (\nabla_q V_{\rm eff})\cdot \nabla_p
\varphi \right) + F\Gamma s \cdot (\nabla_p \varphi - T^{-1} p)  ~.
$$
Since $\varphi$ does not depend on $s$ we get
$$
\eqalign{
p\cdot \nabla_q \varphi - (\nabla_q V_{\rm eff})\cdot \nabla_p \varphi
\,&=\, 0~, \cr
\nabla_{p_1}\varphi \,&=\, T_\L^{-1}p_1~, \cr
\nabla_{p_n}\varphi \,&=\, T_\R^{-1}p_n  ~.
}
\EQ(sys1)
$$
We will show that  this system of linear equations has no solution
unless $T_\L=T_\R$. To see this we consider the system of equations
$$
\eqalign{
p\cdot \nabla_q \varphi - (\nabla_q V_{\rm eff})\cdot \nabla_p \varphi
\,&=\, 0 \cr
\nabla_{p_1}\varphi \,&=\, T_\L^{-1}p_1 ~.
}
\EQ(sys2)
$$
This system has a solution which is given by
$H_{\rm eff}(q,p) / T_\L $.  We claim that this the unique solution
(up to an additive constant) of \equ(sys2).

If this holds true, then the only solution of \equ(sys1) is given
by  $H_{\rm eff}(q,p) / T_\L $ and this is incompatible with the third
equation in \equ(sys1) when $T_\L\not= T_\R$.

Since \equ(sys2) is a linear inhomogeneous equation, it is enough to
show that the only solutions of the homogeneous equation
$$
\eqalign{
p\cdot \nabla_q \varphi - (\nabla_q V_{\rm eff}) \cdot\nabla_p \varphi
\,&=\, 0~, \cr
\nabla_{p_1}\varphi \,&=\, 0 ~,
}
\EQ(sys3)
$$
are the constant functions.
Since $\nabla_{p_1}\varphi \,=\, 0$, $\varphi$ does not depend on
$p_1$, we conclude that the first equation in \equ(sys3) reads
$$
p_1\cdot\nabla_{q_1} \varphi + f_1(q_1, \dots,q_n , p_2, \dots
p_n)\,=\,0 ~,
$$
where $f_1$ does not depend on the variable $p_1$. Thus we see that
$\nabla_{q_1} \varphi=0$ and therefore $\varphi$ does not depend on
the variable $q_1$ either. By the first equation in \equ(sys3) we now get
$$
-\nabla_{q_1} U^{(2)}_1 (q_1-q_2) \cdot\nabla_{p_2} \varphi + f_2
(q_2, \dots ,q_n , p_2, \dots, p_n)\,=\,0 ~,
$$
where $f_2$ does not depend on $p_1$ and $q_1$. By condition {\bf H2}
we see that $\nabla_{p_2} \varphi = 0$ and hence $f$ does not depend
on $p_2$. Iterating the above procedure we find that the only solutions
of \equ(sys3) are the constant functions.
This concludes the proof of \clm(positive1).

\CLAIM Corollary(flux) The stationary state $\mu=\mu_{T_\L,T_\R}$
produces a non-vanishing mean heat flux in the direction from the hotter to
the colder reservoir.
The mean heat flux vanishes only if $T_\L=T_\R$.

\PROOF The entropy production $\sigma$ is given by
$$
\sigma \,=\, \frac{\Phi_\L}{T_\L} +  \frac{\Phi_\R}{T_\R}~,
$$
where $\Phi_\L$ is the energy flow from the left heat
bath to the chain and similarly for $\Phi_\R$. In the stationary state
we have, by \equ(phii), 
$$
\langle \Phi_\L + \Phi_\R \rangle _{\mu}\,=\,0~,
$$
and therefore
$$
\langle \Phi_\L \rangle _{\mu} \,=\, - \langle \Phi_\R \rangle _{\mu}~.
$$
We obtain from \clm(positive1), for $T_\L\ne T_\R$:
$$
0\,>\,\langle \sigma \rangle _{\mu} \,=\, \bigl( \frac{1}{T_\L} -
\frac{1}{T_\R} \bigr) \langle \Phi_\L \rangle _{\mu}  ~.
$$
If, say,  $T_\L >  T_\R$, we get $\langle \Phi_\L \rangle _{\mu} > 0$
and thus energy flows from the hotter to the cooler heat bath.

\LIKEREMARK{Acknowledgments}We would like to thank
G. Ben Arous,
Ch.~Mazza,
H. Spohn,
D. Stroock,
A.-S. Sznitman,
and
L.E. Thomas
for encouragements and helpful discussions, and
V. Zagrebnov for useful questions and comments.
This work was supported in part by the
Fonds National Suisse.

\SECTIONNONR References

\eightpoint{
\setitemindent{[CELS]}\refindent=\itemindent

\ref
\no CELS
\by N.I. Chernov, G.L. Eyink, J.L. Lebowitz, and Ya. G. Sinai
\paper Steady-state electric conduction in the periodic Lorentz gas
\jour Commun. Math. Phys.
\vol 154
\pages 569--601
\yr 1993
\endref

\ref
\no CL
\by N.I. Chernov and J.L. Lebowitz 
\paper Stationary nonequilibrium states in boundary driven Hamiltonian
systems: Shear flow
\preprint 
\yr 1997 
\endref

\ref
\no ECM
\by D.J. Evans, E.G.D. Cohen, and G.P. Morriss
\paper Probability of second Law violations in shearing steady flows
\jour Phys. Rev. Lett.
\vol 71
\pages 2401--2404
\yr 1993
\endref

\ref
\no EM
\by D. Evans and G. Morriss
\book Statistical Mechanics of
Nonequilibrium Liquids
\publisher New York, Academic Press
\yr 1990
\endref

\ref
\no EPR
\by J.-P. Eckmann, C.-A. Pillet, and L. Rey-Bellet
\paper Non-equilibrium statistical mechanics of 
anharmonic chains coupled to two heat baths at different temperatures.
Commun.Math.Phys., in print
\endref

\ref
\no FGS
\by J. Farmer, S. Goldstein, and E.R. Speer
\paper Invariant states of a thermally conducting barrier
\jour J. Stat. Phys.
\vol 34
\pages 263--277
\yr 1984
\endref

\ref
\no G
\by G. Gallavotti
\paper Chaotic hypothesis and universal
large deviations properties
\jour Doc. Math. J. DMV Extra Volume ICM I 
\pages 65--93
\yr 1998
\endref

\ref
\no GC1
\by  G. Gallavotti and E.G.D. Cohen
\paper Dynamical ensembles in nonequilibrium statistical mechanics
\jour Phys. Rev. Lett.
\vol 74 \pages 2694--2697
\yr 1995
\endref

\ref
\no GC2
\by  G. Gallavotti and E.G.D. Cohen
\paper Dynamical ensembles in stationary states
\jour J. Stat. Phys,
\vol 80 \pages 931--970
\yr 1995
\endref
\ref
\no GKI
\by
S.  Goldstein, C. Kipnis, and N. Ianiro
\paper Stationary states for a mechanical system with stochastic boundary
conditions. 
\jour J. Stat. Phys. 
\vol 41  
\pages 915--939 
\yr 1985
\endref

\ref
\no GLP
\by S. Goldstein, J.L. Lebowitz, and E. Presutti
\paper Stationary states for a mechanical system with stochastic boundaries
\inbook {\it Random Fields}, (Colloquia Mathematicae Societatis J\'anos
Bolyai 
\vol 27
\publisher Amsterdam, North-Holland 
\yr 1981
\endref

\ref
\no H
\by W.G. Hoover
\book Computational Statistical Mechanics
\publisher Elsevier
\yr 1991
\endref

\ref
\no IK
\by K. Ishihara and H. Kunita
\paper A classification of the second order degenerate elliptic operators
and its probabilistic characterization
\jour Z. Wahrsch. und Verw. Geb.
\vol 39
\pages 235--254
\yr 1974
\endref

\ref
\no K
\by J. Kurchan
\paper Fluctuation theorem for stochastic dynamics 
\jour J. Phys.
\vol A31
\pages 3719--3729 
\yr 1998
\endref

\ref
\no Ku
\by H. Kunita
\paper Supports of diffusion processes and controllability problems
\inbook Proc. Intern. Symp. SDE Kyoto
\pages 163--185
\yr 1976
\endref

\ref
\no LS
\by J.L. Lebowitz and H. Spohn
\paper The Gallavotti-Cohen fluctuation theorem for stochastic
dynamics
\preprint 
\yr 1998
\endref

\ref
\no PH
\by H.A. Posch and W.G. Hoover
\jour Phys. Rev.
\paper Equilibrium and non equilibrium Lyapunov spectra for dense
fluids and solids
\vol A39
\pages 2175--2188
\yr 1989
\endref

\ref
\no SL
\by H. Spohn and J.L. Lebowitz, Stationary non-equilibrium
\paper states of infinite harmonic systems
\jour Commun. Math. Phys.
\vol 54
\pages 97
\yr 1977
\endref

\ref
\no SV
\by D.W. Stroock and S.R.S. Varadhan
\paper On the support of diffusion processes with applications to the
strong maximum principle
\inbook Proc. 6-th Berkeley Symp. Math. Stat. Prob. 
\vol III
\pages 333--368
\yr 1972
\endref
}

\bye